\documentclass[journal]{IEEEtran}

\usepackage{float}  

\usepackage{caption}
\usepackage{subcaption}

\usepackage{stfloats}

\usepackage{algorithm}  
\usepackage{algpseudocode}  
\usepackage{amsmath}  

\usepackage{graphicx}  
\usepackage{color}
\usepackage{cite}

\usepackage{graphicx}  
\usepackage{epstopdf}

\usepackage{graphicx}  

\usepackage{multirow,booktabs} 
\usepackage{makecell} 
\usepackage{diagbox}

\usepackage{stfloats}
\usepackage{graphicx}

\usepackage{url}
\usepackage[hidelinks]{hyperref}

\ifCLASSINFOpdf
\else
\fi
\usepackage{amsthm,amsmath,amssymb}

\begin{document}
	
\captionsetup[figure]{labelfont={},labelformat={default},labelsep=period,name={Fig.}}


\captionsetup[table]{labelfont={},labelformat={default},labelsep=period,name={Table}}

%
\title{U-Net-Based Generative Joint Source-Channel Coding for Wireless Image Transmission}







\author{Ming~Ye,~\IEEEmembership{Member,~IEEE,}~Kui Cai,~\IEEEmembership{Senior Member,~IEEE,}~Cunhua Pan,~\IEEEmembership{Senior Member,~IEEE,} Zhen Mei,~\IEEEmembership{Member,~IEEE,} Wanting Yang, Chunguo Li,~\IEEEmembership{Senior Member,~IEEE}, and Bo Wang,~\IEEEmembership{Senior Member,~IEEE}

\thanks{This work was supported in part by the Joint SMU-SUTD Grant 24-SIS-SMU-052, in part by A*STAR under its RIE2025 Innovation \& Enterprise Industry Alignment Fund -- Industry Collaboration Projects (Award I2501E0064, Grant H25-MCP3438), in part by the National Natural Science Foundation of China under Grants U25A20393, U25B2007, and 62331023, and in part by the National Key R\&D Program of China under Grant 2024YFC3807900.

M. Ye, K. Cai, W. Yang, and B. Wang are with the Singapore University of Technology and Design, Singapore 487372 (e-mail: 230198460@aa.seu.edu.cn; cai$_{-}$kui@sutd.edu.sg; wanting$_{-}$yang@sutd.edu.sg; bo$_{-}$wang@sutd.edu.sg).	
	
C. Pan and C. Li are with the National Mobile Communications Research Laboratory, Southeast University, Nanjing 210096, China (e-mail: cpan@seu.edu.cn;  chunguoli@seu.edu.cn).

Z. Mei is with
the School of Electronic and Optical Engineering, Nanjing University of Science and Technology, Nanjing 210094, China (e-mail: meizhen@njust.edu.cn). \textcolor{blue}{The source code is publicly available at:} https://github.com/Yeming198460/U-Net-Based-Generative-Joint-Source-Channel-Coding.

}

}
 
 


%
%

\markboth{}%
{Shell \MakeLowercase{\textit{et al.}}: Bare Demo of IEEEtran.cls for IEEE Journals}
%



\maketitle

\begin{abstract}

Deep learning (DL)-based joint source-channel coding (JSCC) methods have achieved remarkable success in wireless image transmission. However, these methods either focus on conventional distortion metrics that do not necessarily yield high perceptual quality or incur high computational complexity.  In this paper, we propose two DL-based JSCC (DeepJSCC) methods that leverage deep generative architectures for wireless image transmission.
Specifically, we propose G-UNet-JSCC, a scheme comprising an encoder and a U-Net-based generator serving as the decoder. 
Its skip connections enable multi-scale feature fusion by integrating low- and high-level features, thereby improving both the pixel-level fidelity and perceptual quality of reconstructed images.
To further enhance pixel-level fidelity, the encoder and the U-Net-based decoder are jointly optimized using a weighted sum of structural similarity and mean-squared error (MSE) losses.
Building upon G-UNet-JSCC, we further develop a DeepJSCC method called cGAN-JSCC, where the decoder is enhanced through adversarial training. In this scheme, we retain the encoder of G-UNet-JSCC and adversarially train the decoder's generator against a patch-based discriminator. cGAN-JSCC employs a two-stage training procedure. The outer stage trains the encoder and the decoder end-to-end using the same weighted sum of these losses, while the inner stage adversarially trains the decoder's generator and the discriminator by minimizing a joint loss combining adversarial and distortion losses. 
Simulation results demonstrate that the proposed methods  achieve superior pixel-level fidelity and perceptual quality compared with existing DeepJSCC methods for both low- and high-resolution images.
cGAN-JSCC achieves better reconstruction performance than G-UNet-JSCC.


\end{abstract}



%
\IEEEpeerreviewmaketitle

\section{Introduction}
%
%
%
%

\vspace{-0.5mm}  

\IEEEPARstart{T}{he}  
proliferation of wireless multimedia applications, such as virtual reality and augmented reality, has imposed stringent low-latency and high-efficiency requirements on image/video transmission. To achieve reliable data transfer, wireless communication systems are carefully designed and optimized. In conventional communication systems following Shannon's separation theorem \cite{b1}, the transmitter comprises separate source and channel encoders. The source encoder compresses data by removing redundancy, whereas the channel encoder introduces controlled redundancy to protect against channel impairments. At the receiver, a corresponding pair of channel and source decoders perform the inverse operations to reconstruct the original data.

Shannon established that the separation of source and channel coding is theoretically optimal in the asymptotic regime of infinite blocklength. 
However, the separation-based approach becomes highly suboptimal under the practical finite blocklength regime, especially for communication systems operating with strict latency and bandwidth constraints \cite{bb2, bb3}. 
Specifically, the substantial time required for compressing and decompressing high-rate image/video sources has become a primary bottleneck in end-to-end latency. Consequently, joint source-channel coding (JSCC), which unifies source and channel coding within a single architecture, is positioned as a viable solution for mission-critical 6G applications that require ultra-low latency, including autonomous vehicles and remote surgery \cite{bb4}. 
Despite the established benefits of JSCC, designing effective JSCC approaches remains a long-standing challenge. Most research on traditional JSCC is confined to theoretical scenarios \cite{b2, b3} or focuses on jointly optimizing parameters of separately designed source and channel coding \cite{b4, b5, b6}. This limitation stems both from a scarcity of practical JSCC methods and from the lack of theoretical performance bounds on the achievable mean squared-error (MSE) distortion for lossy JSCC at finite blocklengths \cite{b7}, particularly for practical sources like images.

Recently, the introduction of deep neural networks (DNNs) has rapidly advanced deep learning (DL)-based JSCC (DeepJSCC) methods. These DNN-based schemes establish an end-to-end learning framework, enabling a direct mapping from source signals to continuous channel inputs at the encoder and the inverse mapping at the decoder.
Such end-to-end learning is enabled the powerful ability of DNNs to extract inherent features and implicitly learn to compensate for channel impairments \cite{b8,b9,b10}.  
For example, the convolutional neural network (CNN)-based DeepJSCC method proposed in \cite{b11} is a seminal work for wireless image transmission, as it was the first to operate without explicit source compression or channel coding. It demonstrates superior performance over traditional separation-based systems that combine an explicit source codec (e.g., JPEG) with a channel code such as a low-density parity-check (LDPC) code.
The initial DeepJSCC framework was subsequently extended to handle multi-path fading channels \cite{b13} and support adaptive bandwidth \cite{b14,b15} and rate control \cite{b16,b17} for wireless image transmission.
To reduce model complexity, the authors of \cite{ba17} proposed DeepJSCC-T, a lightweight DeepJSCC scheme that uses ConvNeXt-based modules for wireless image transmission.
Following these developments, the authors of \cite{b18, b19} extended the DeepJSCC approach to multiple-input multiple-output (MIMO) systems by exploiting the self-attention mechanism of the Vision Transformer.
To further reduce the computational complexity of Transformer-based DeepJSCC frameworks, the authors of \cite{ba19} proposed SwinJSCC, which adopts a Swin Transformer backbone.
Research on DeepJSCC methods has also explored security in semantic communication systems,
employing both conventional encryption methods \cite{b20} and encryption schemes based on neural networks \cite{b21, b22}. 
Although training DeepJSCC schemes can be computationally intensive, the inference of DeepJSCC schemes is generally more efficient than that of traditional separation-based systems.
This is because DeepJSCC schemes perform end-to-end encoding and decoding directly, whereas traditional systems typically require advanced image codecs followed by capacity-approaching channel codecs. 
However, the aforementioned DeepJSCC approaches still face two major limitations. First, they primarily focus on conventional reconstruction quality metrics, such as peak signal-to-noise ratio (PSNR) and structural similarity index measure (SSIM), without explicitly considering the perceptual similarity between the source and reconstructed images.
Second, some DeepJSCC methods use sophisticated network architectures, such as Transformer-based architectures, to achieve competitive reconstruction performance but can incur relatively high model complexity, limiting their deployment on resource-constrained edge devices.



 


Inspired by the success of deep generative models (DGMs) in image generation \cite{b23, b24, b25}, researchers have proposed DeepJSCC methods based on DGMs for wireless image transmission. These methods leverage DGMs such as diffusion models \cite{b26}, variational autoencoders (VAEs) \cite{b27, b28}, and generative adversarial networks (GANs) \cite{b29, bb12}. 
Specifically, a three-stage DeepJSCC approach based on diffusion models was proposed in \cite{b26} for wireless image transmission in semantic communication systems.     
The authors of \cite{b27} proposed a VAE-based DeepJSCC scheme for distributed Gaussian sources over a multiple-access additive white Gaussian noise (AWGN) channel, exploiting a key regularization mechanism to enhance reconstruction quality.  
In \cite{b28}, a DeepJSCC method was proposed for analog noise channels by using the VAE loss function to characterize the rate-distortion trade-off. Its efficacy was demonstrated through simulations involving Gaussian sources over AWGN channels.
By adversarially training a legitimate autoencoder against an adversary network, the DeepJSCC method proposed in \cite{b29} achieved a privacy-distortion trade-off for secure image transmission over a noisy wiretap channel. A DeepJSCC scheme was proposed in \cite{bb12} to enhance the reconstruction quality of images over multipath fading channels by adversarially training a decoder network with a discriminator.
However, these DeepJSCC methods \cite{b26, b27, b28, b29, bb12} either fail to incorporate  the learned perceptual image patch similarity (LPIPS) metric \cite{b31} that better aligns with human perception than conventional metrics or incur high computational complexity.
Specifically, among GAN-related JSCC methods, the generator architecture in \cite{bb12} constrains perceptual quality, as it lacks efficient structures such as the StyleGAN-2 generator employed in \cite{b32}, thereby limiting its capacity for high-quality image reconstruction. 
In addition, the joint optimization mechanism between the adversarially trained decoder and the encoder in \cite{bb12} is not clearly described.
The authors of \cite{b30} proposed the first DeepJSCC scheme that leverages a pre-trained StyleGAN-2 generator to improve both pixel-level fidelity and perceptual quality of reconstructed images for wireless image transmission. However, this approach is not only confined to face images but also incurs high computational complexity, as both its encoder and decoder employ complex modules like attention mechanisms and residual blocks.





In our prior works \cite{b34, b35}, GAN variants leveraging a generator based on the U-Net architecture in \cite{b25, b36} have been successfully applied to channel estimation by formulating it as an image-to-image translation task.  
Inspired by this success, we propose two DeepJSCC methods for wireless image transmission. These methods are based on deep generative architectures, specifically a conditional GAN (cGAN) framework with U-Net-based generators. In particular, we propose G-UNet-JSCC, which comprises an encoder and a U-Net-based generator serving as the decoder.  
In the U-Net-based decoder, the skip connections  enable multi-scale information fusion by combining low-level features (e.g., textures and edges) with high-level features (e.g., overall structure). This fusion improves both the pixel-level fidelity (e.g., PSNR and SSIM) and the perceptual quality (e.g., LPIPS), particularly under low signal-to-noise ratio (SNR) conditions. 
Compared to prior DeepJSCC schemes that employ almost symmetric encoder-decoder structures (e.g., \cite{b11, bb12}) or complex encoders (e.g., \cite{b18, b30}), our G-UNet-JSCC adopts an asymmetric design: a relatively simple encoder paired with a powerful U-Net decoder.
This design enables high reconstruction performance while maintaining relatively low model complexity compared with other high-performing DeepJSCC methods.
To further improve the pixel-level fidelity, we train the encoder and the decoder end-to-end by minimizing a weighted sum of MSE and SSIM losses.     
Building upon G-UNet-JSCC, we then propose a DeepJSCC method,  named cGAN-JSCC, which enhances the decoder of G-UNet-JSCC via adversarial training. 
In cGAN-JSCC, we retain the G-UNet-JSCC encoder and adversarially train the decoder's U-Net-based generator against a patch-based discriminator.    
cGAN-JSCC employs a two-stage training procedure. In the outer stage, the encoder and the decoder are trained end-to-end using the weighted sum of MSE and SSIM losses. In the inner stage, the generator and the discriminator are adversarially trained by 
minimizing a joint loss that combines adversarial and L2 losses \cite{b25}.
This two-stage training strategy achieves high reconstruction performance through optimization in the outer stage and further improves reconstruction quality through adversarial training in the inner stage.
Extensive numerical experiments  demonstrate that both the proposed methods achieve high reconstruction quality on high- and low-resolution images. For low-resolution images, cGAN-JSCC outperforms G-UNet-JSCC in terms of reconstruction performance and robustness.




The main contributions of this work can be summarized as follows:

$\bullet$  We employ a U-Net-based generator as the decoder in our JSCC frameworks. Its skip connections facilitate multi-scale feature fusion, integrating low-level details with high-level semantics. 
This design improves reconstruction quality across different SNR conditions, while maintaining relatively low model complexity.
To our knowledge, this work is among the first to develop U-Net-based JSCC methods for wireless image transmission, achieving both high pixel-level fidelity and perceptual quality.






$\bullet$ We first propose G-UNet-JSCC, a DeepJSCC scheme for wireless image transmission. Unlike existing DL-based encoder-decoder structures, G-UNet-JSCC adopts an asymmetric architecture, pairing a simple encoder with a powerful U-Net-based decoder. This asymmetric design improves both the pixel-level fidelity and the perceptual quality, especially in low-SNR regimes. To further enhance the pixel-level fidelity, the encoder and the decoder are jointly optimized to minimize a weighted sum of the MSE and SSIM losses.



$\bullet$ We then propose cGAN-JSCC, which incorporates adversarial training into the U-Net-based JSCC decoder. In cGAN-JSCC, we keep the same encoder as G-UNet-JSCC and adversarially train the U-Net-based generator against a patch-based discriminator within the decoder. 
The cGAN-JSCC training follows a two-stage procedure: an outer stage for end-to-end optimization of the encoder and the decoder, and an inner stage within the decoder for adversarial training of the generator and the discriminator.  This strategy further improves reconstruction quality through adversarial training.

$\bullet$ 
Extensive numerical experiments on multiple datasets demonstrate the superiority of cGAN-JSCC and G-UNet-JSCC in terms of pixel-level fidelity and perceptual quality for both low- and high-resolution images. Moreover, cGAN-JSCC further improves the reconstruction performance of G-UNet-JSCC.

\section{System  Model}

\begin{figure}
	\centering
	\includegraphics[width=7.5cm]{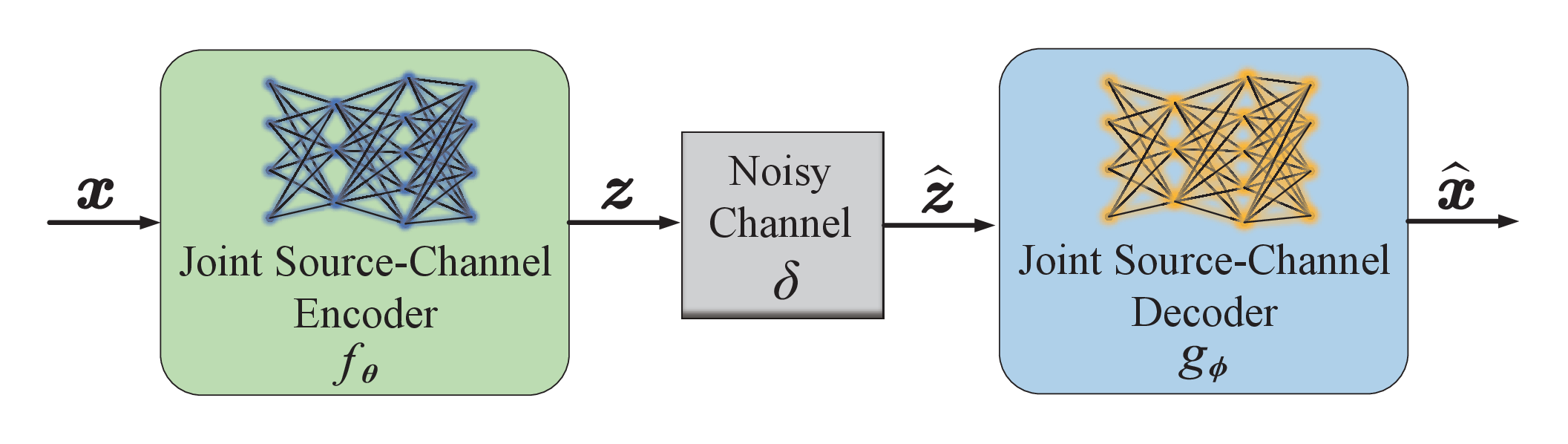}
	\caption{Block diagram of the image transmission system for the proposed G-UNet-JSCC method.}
	\label{jscc_image_transmission}
\end{figure}

We consider a communication system in which a source signal is transmitted from a transmitter to a receiver over a noisy wireless channel.
As shown in Fig. \ref{jscc_image_transmission}, the joint source-channel encoder in the G-UNet-JSCC method directly maps a real-valued source input (image) $\boldsymbol{x}\in \mathbb{R} ^n$ into a complex-valued channel input symbol vector $\boldsymbol{z}\in \mathbb{C} ^k$ for channel transmission. The encoded source image can be expressed as 
\begin{equation}
\boldsymbol{z}=f_{\boldsymbol{\theta}}\left( \boldsymbol{x} \right) \in \mathbb{C} ^k,
\label{encoding}	
\end{equation}
where $f_{\boldsymbol{\theta}}: \mathbb{R}^n \rightarrow \mathbb{C}^k$ denotes the parameterized encoding function with parameters $\boldsymbol{\theta}$. Here, $n$ and $k$ represent the source dimension and the channel dimension, respectively. The bandwidth compression ratio (BCR) is defined as $r = k/n$. A smaller $r$ corresponds to a higher level of compression.



The output of the final convolutional layer of the encoder is transformed into a vector of complex-valued symbols. This vector, denoted as $\tilde{\boldsymbol{z}}$, is then normalized before being transmitted through the noisy wireless channel. 
To comply with the average transmit power constraint $\bar{P}$, the power normalization is performed according to:
\begin{equation}
\boldsymbol{z}=\sqrt{k\bar{P}}\frac{\tilde{\boldsymbol{z}}}{\sqrt{\tilde{\boldsymbol{z}}^H\tilde{\boldsymbol{z}}}},
\label{power_constraint}
\end{equation}
where $\boldsymbol{z}$ is the normalized feature vector for transmission.




After encoding, the normalized feature vector $\boldsymbol{z}$ is transmitted over a Rayleigh slow-fading channel, resulting in a received signal $\hat{\boldsymbol{z}}$. This transmission process can be modeled by a random corruption function $\delta:\mathbb{C}^k\rightarrow\mathbb{C}^k$. Specifically, the channel input-output relationship is given by
\begin{equation}
	\hat{\boldsymbol{z}}
	=
	\delta(\boldsymbol{z},\sigma^2)
	=
	h\boldsymbol{z}
	+
	\boldsymbol{n},
	\label{transfer_function}
\end{equation}
where $h\sim\mathcal{CN}(0,1)$ denotes the complex channel coefficient of the Rayleigh fading channel, and $\boldsymbol{n}\sim\mathcal{CN}(\mathbf{0},\sigma^2\mathbf{I})$ denotes complex AWGN, with $\sigma^2$ representing the noise power. 
Other channel models can also be integrated into the end-to-end system, provided that the corresponding corruption function is differentiable \cite{b11, b18}. For example, the corruption function of an AWGN channel is expressed as $\hat{\boldsymbol{z}}=\delta(\boldsymbol{z},\sigma^2)=\boldsymbol{z}+\boldsymbol{n}$.




The received signal $\hat{\boldsymbol{z}}$ is then mapped to the reconstructed image $\hat{\boldsymbol{x}}$ by the joint source-channel decoder using a decoding function $g_{\boldsymbol{\phi}}:\mathbb{C}^k\rightarrow\mathbb{R}^n$, which is parameterized by $\boldsymbol{\phi}$. The reconstructed image $\hat{\boldsymbol{x}}$ can be expressed as
\begin{equation}
	\hat{\boldsymbol{x}}=g_{\boldsymbol{\phi}}\left( \hat{\boldsymbol{z}} \right) \in \mathbb{R} ^n.
	\label{decoding}
\end{equation}

The objective of the end-to-end system is to jointly optimize the encoder and decoder parameters $\boldsymbol{\theta}$ and $\boldsymbol{\phi}$ to minimize the average distortion between the source image $\boldsymbol{x}$ and its reconstruction $\hat{\boldsymbol{x}}$:
\begin{equation}
\left( \boldsymbol{\theta }^*, \boldsymbol{\phi }^* \right) =\mathrm{arg}~\underset{\boldsymbol{\theta },\boldsymbol{\phi }}{\min}~\mathbb{E} _{p\left( \boldsymbol{x},\hat{\boldsymbol{x}} \right)}\left[ d\left( \boldsymbol{x},\hat{\boldsymbol{x}} \right) \right],
\end{equation}
where $p(\boldsymbol{x}, \hat{\boldsymbol{x}})$ denotes the induced joint probability distribution of $\boldsymbol{x}$ and $\hat{\boldsymbol{x}}$, and $d(\boldsymbol{x}, \hat{\boldsymbol{x}})$ is an arbitrary distortion measure.
Conventional metrics such as PSNR and SSIM do not adequately capture perceptual quality. Therefore, this work aims to improve both the pixel-wise fidelity and the perceptual quality of the reconstructed images, particularly under low-SNR conditions.


\begin{figure*}
	\centering
	\includegraphics[width=15.0cm]{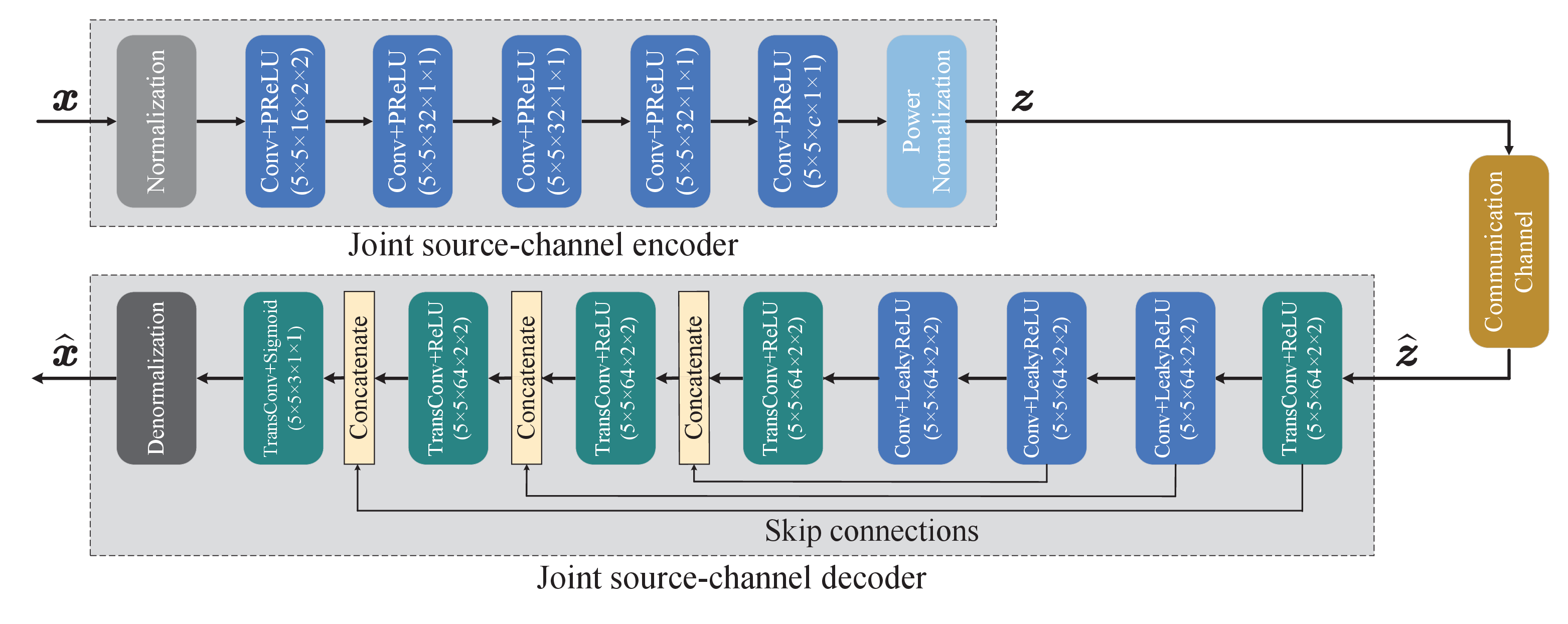}
	\caption{Encoder and decoder DNN architectures of the proposed G-UNet-JSCC method.}
	\label{fig.2_DeepJSCC_generator}
\end{figure*}

\section{Proposed Methods}

In this section, we propose two DeepJSCC methods based on deep generative architectures to improve both the pixel-level fidelity and the perceptual quality of reconstructed images for wireless image transmission. 
We first propose G-UNet-JSCC, a DeepJSCC method that comprises an encoder and a U-Net-based generator serving as the decoder. 
The skip connections in the U-Net-based decoder facilitate multi-scale feature fusion by combining low-level features  with high-level features, thereby improving both the pixel-level fidelity and the perceptual quality.
To further improve the pixel-level fidelity of reconstructed images, the encoder and the U-Net-based decoder are optimized jointly to minimize a weighted sum of MSE and SSIM losses.
Furthermore, we propose cGAN-JSCC, an adversarially trained variant of G-UNet-JSCC. 
In cGAN-JSCC, we keep the same encoder as G-UNet-JSCC and adversarially train the generator against a patch-based discriminator within the decoder. 
The cGAN-JSCC method is trained through a two-stage procedure. The encoder and the U-Net-based generator are jointly optimized using a combined loss function in the outer training stage, while the U-Net-based generator and the discriminator are trained with a combination of adversarial and L2 losses in the inner training stage.

\subsection{ G-UNet-JSCC Framework}

To enhance both the pixel-wise accuracy and the perceptual quality of reconstructed images, we propose a DeepJSCC framework, termed G-UNet-JSCC. The G-UNet-JSCC method consists of a deep encoder and a U-Net-based generator serving as the decoder. This end-to-end scheme jointly optimizes the encoding and decoding functions, which are parameterized by DNNs with learnable parameters $\boldsymbol{\theta}$ and $\boldsymbol{\phi}$, respectively.

As illustrated in Fig. \ref{fig.2_DeepJSCC_generator}, the source image $\boldsymbol{x}$ is processed by the encoder in the G-UNet-JSCC method and normalized according to Eq. \eqref{power_constraint} at the final layer of the encoder to comply with the power constraint. Subsequently, the encoded signal $\boldsymbol{z}$ is transmitted over the wireless noisy channel  characterized by the transfer function defined in Eq. \eqref{transfer_function}. 
At the receiver, the corrupted signal $\hat{\boldsymbol{z}}$ is decoded by the U-Net-based decoder to reconstruct the source image:
\begin{equation}
	\hat{\boldsymbol{x}} = g_{\boldsymbol{\phi}}\left( \delta \left( f_{\boldsymbol{\theta}}\left( \boldsymbol{x} \right), \sigma^2 \right) \right).
\end{equation}

Inspired by the combined loss function in \cite{b18}, which consists of an MSE loss and an LPIPS loss, we replace the LPIPS loss with an SSIM loss to reduce the training overhead. Unlike the LPIPS loss that requires an additional pretrained feature extraction network, the SSIM loss directly evaluates the structural similarity between the original and reconstructed images. Thus, the encoder and decoder in the G-UNet-JSCC scheme are optimized end-to-end by minimizing the following combined loss function:
\begin{equation}
	\mathcal{L}_{\mathrm{C}}(\boldsymbol{\theta}, \boldsymbol{\phi}) =  \mathcal{L}_{\mathrm{MSE}}(\boldsymbol{x}, \hat{\boldsymbol{x}}) + \lambda_{\mathrm{S}} \mathcal{L}_{\mathrm{SSIM}}(\boldsymbol{x}, \hat{\boldsymbol{x}}),
	\label{G-UNet-jscc_combined_loss}
\end{equation}
where $\mathcal{L}_{\mathrm{MSE}}$ and $\mathcal{L}_{\mathrm{SSIM}}$ denote the MSE and SSIM loss functions, respectively, and $\lambda_{\mathrm{S}}$ controls the relative importance of the SSIM loss. In our implementation, $\lambda_{\mathrm{S}}=0.1$. The combined loss function is a weighted sum of the MSE and SSIM losses. 
The objective of training the G-UNet-JSCC model is to find the optimal parameters $\left( \boldsymbol{\theta }^*, \boldsymbol{\phi }^* \right)$ that minimize the combined loss:
$\left( \boldsymbol{\theta }^*, \boldsymbol{\phi }^* \right) =\mathrm{arg}~ \underset{\boldsymbol{\theta },\boldsymbol{\phi }}{\min}~\mathbb{E} \left[ \mathcal{L} _{\mathrm{C}}\left( \boldsymbol{\theta },\boldsymbol{\phi }\right)  \right]$.


To achieve a favorable trade-off between computational complexity and reconstruction quality, we employ a DNN architecture comprising an encoder and a U-Net-based decoder, as depicted in Fig. \ref{fig.2_DeepJSCC_generator}. The U-Net-based decoder, with its efficient skip connections, is particularly effective for image reconstruction tasks. As a concrete example to illustrate the architecture of G-UNet-JSCC, we use images of dimensions $256 \times 256 \times 3$ (i.e., height, width, and color channels) cropped from the Celeb-HQ dataset \cite{b30, b37}.
At the encoder, the source image $\boldsymbol{x}$ is first processed by a normalization layer. 
The normalization layer scales the input pixel values of the source image from [0, 255] to [0, 1] through division by the maximum pixel value 255 for 8-bit images, which helps mitigate potential gradient explosion.
The normalized image is then processed by five convolutional layers. 
The first convolutional layer performs downsampling to extract low-level features. The subsequent four convolutional layers, all operating at a fixed resolution, conduct non-linear transformations to synthesize features, with the final convolutional layer producing $c$ feature maps of dimensions $128 \times 128$.
All convolutional layers in the encoder use the parametric rectified linear unit (PReLU) as their activation functions. 
To satisfy the power constraint, the final convolutional layer at the encoder, which comprises $c$ output channels, is followed by a power normalization layer. 
The number of output channels $c$ can be adjusted to control the BCR.  
Note that 
the real-valued output from the encoder's final convolutional layer is transformed into a vector of $k$ complex-valued channel input symbols before the power normalization layer.
This power normalization layer, as defined in Eq. \eqref{power_constraint}, produces the normalized feature vector $\boldsymbol{z}$ for transmission. The specific configurations of these hyperparameters used in our experiments are provided in Fig. \ref{fig.2_DeepJSCC_generator}. In Fig. \ref{fig.2_DeepJSCC_generator}, the notation $F \times F \times K \times S_{\mathrm{h}} \times S_{\mathrm{w}}$ defines a (transposed) convolutional layer, where $K$ denotes the number of filters with a spatial size of $F \times F$, and $S_{\mathrm{h}}\times S_{\mathrm{w}}$ represents the stride in the height and width dimensions, respectively.


\begin{algorithm}
	\caption{G-UNet-JSCC Joint Training Algorithm}	
 
	\begin{algorithmic}[1]
\State \textbf{Input:} Training dataset $\mathcal{D} = \{\boldsymbol{x}_i\}_{i=1}^T$, the number of epochs $E$, BCR set $R$, batch size $B$, SNR set $S$, learning rate $\eta$


\State \textbf{Output:} Trained encoders $f_{\boldsymbol{\theta}}$ and decoders $g_{\boldsymbol{\phi}}$ for all $(r,\gamma)$ combinations


		\For{each bandwidth ratio $r \in R$}	
\For{each SNR value $\gamma \in S$}

\State Initialize encoder $f_{\boldsymbol{\theta }}$ and decoder $g_{\boldsymbol{\phi }}$

\State Compute noise variance: $\sigma^2 \gets 10^{-\gamma/10}$	
 
		\For{epoch $= 1$ to $E$}
		
 
\State \textbf{\textit{Forward pass through JSCC pipeline}}

 		
\State Sample a minibatch of size $B$ from $\mathcal{D}$		
		

		\State Encode the source image $\boldsymbol{x}$ to the complex-valued symbol vector $\boldsymbol{z}$ using Eq.~\eqref{encoding}: $\boldsymbol{z}\gets f_{\boldsymbol{\theta }}(\boldsymbol{x})$

		
\State Add the AWGN to the vector $\boldsymbol{z}$ using Eq. \eqref{transfer_function} and obtain the corrupted signal $\hat{\boldsymbol{z}}$ 

\State Decode to reconstruct the image $\hat{\boldsymbol{x}}$ from the corrupted signal $\hat{\boldsymbol{z}}$: $\hat{\boldsymbol{x}} \gets g_{\boldsymbol{\phi}}(\hat{\boldsymbol{z}})$
		
\State \textbf{\textit{Compute joint reconstruction loss}}

\State Compute reconstruction loss $\mathcal{L} _{\mathrm{C}}$ using Eq.~\eqref{G-UNet-jscc_combined_loss}	
		
		
\State \textbf{\textit{Joint parameter update via backpropagation}}


\State Update parameters $\{\boldsymbol{\theta}, \boldsymbol{\phi}\}$ by gradient descent on $\mathcal{L}_{\mathrm{C}}$: $\boldsymbol{\theta}, \boldsymbol{\phi} \leftarrow \boldsymbol{\theta}, \boldsymbol{\phi} - \eta \nabla_{\boldsymbol{\theta}, \boldsymbol{\phi}} \mathcal{L}_{\mathrm{C}}$

		
		\EndFor

\State Save trained models for current $(r, \gamma)$ configuration: encoder $f_{\boldsymbol{\theta}}$ and decoder $g_{\boldsymbol{\phi}}$		
		
		\EndFor
		\EndFor
		

		
	\end{algorithmic}
	\label{G-UNet-jscc_training1}	
\end{algorithm}

The corrupted signal $\hat{\boldsymbol{z}}$, a vector of $k$ complex-valued channel output symbols, is converted into a vector of $2k$ real-valued symbols by concatenating its real and imaginary components.  
The real-valued vector is then reshaped into a real-valued channel output of dimensions $128 \times 128 \times c$, which serves  as the input of the U-Net-based decoder.
After that, the channel output is first upsampled by a transposed convolutional layer, yielding 64 feature maps of dimensions $256 \times 256$. 
Subsequently, these feature maps are processed and downsampled  by three convolutional layers to generate 64 feature maps of dimensions $32 \times 32$. Following this, the resulting feature maps are progressively upsampled by the next three transposed convolutional layers to produce 64 feature maps of size $256 \times 256$. 
Then, the last transposed convolutional layer reduces the channel dimension from 64 to 3, producing a $256 \times 256 \times 3$ tensor.
Finally, a denormalization layer is applied to scale the pixel values from the output of the decoder's final transposed convolutional layer to the [0, 255] range, yielding the final reconstructed image $\hat{\boldsymbol{x}}$ of size $256 \times 256 \times 3$.
In the decoder, the transposed convolutional layers use the rectified linear unit (ReLU) activation function, except for the final transposed convolutional layer, which employs the Sigmoid activation function. The convolutional layers in the decoder use the leaky ReLU (LeakyReLU) activation function.

The U-Net-based decoder incorporates its own downsampling-upsampling path
to facilitate multi-scale feature fusion via internal skip connections. These connections integrate the upsampled feature maps with the feature maps from other parts of the decoder at the corresponding scale.
For example, as shown in Fig. \ref{fig.2_DeepJSCC_generator}, the $256 \times 256$ feature maps from the decoder's fourth transposed convolutional layer are combined with those from the decoder's first transposed convolutional layer via skip connections. Similarly, the $64 \times 64$ feature maps from the second transposed convolutional layer are fused with those from the decoder's second convolutional layer.  
Despite being corrupted by channel noise, the output from the encoder that serves as the input to the U-Net-based decoder retains rich low-level information such as textures and edges. 
This information provides valuable spatial cues that help the decoder preserve fine-grained details and maintain consistency with the transmitted information, thereby promoting pixel-wise fidelity.
The skip connections supply corrupted low-level information, while the backbone path provides corrupted high-level information. 
Although both the low-level and high-level information are affected by channel noise, their content is complementary.
Without these skip connections, the decoder is prone to generating structurally distorted images.
Moreover, by leveraging low-level information from skip connections, high-level representations from the deeper layers of the backbone path, and image priors learned during training, the U-Net-based decoder effectively restores image regions corrupted by channel noise.
This results in clearer and more natural textures and enhances the decoder's ability to reconstruct images under channel noise, which is important for improving perceptual quality.


The training procedure of the G-UNet-JSCC scheme is illustrated in Algorithm \ref{G-UNet-jscc_training1}. 
Although the encoder and the U-Net-based decoder are end-to-end trained, they can be deployed separately after training. 
Compared to existing DL-based encoder-decoder structures such as almost symmetric structures or complex encoder structures, our G-UNet-JSCC adopts a deliberately asymmetric architecture. It pairs a relatively simple encoder with a powerful U-Net-based decoder.  Hence, the G-UNet-JSCC scheme is particularly suitable for practical scenarios where the encoder is resource-constrained (e.g., on mobile devices), while the decoder can leverage more computational resources at the receiver.
To maximize perceptual quality and minimize distortion, our G-UNet-JSCC employs a U-Net-based generator as its decoder. 
This design is inspired by the success of deep generative architectures like cGANs, which often use U-Net structures in their generators for high-fidelity image synthesis.




\subsection{cGAN-JSCC Framework}

 \begin{figure*}
	\centering
	\includegraphics[width=14.3cm]{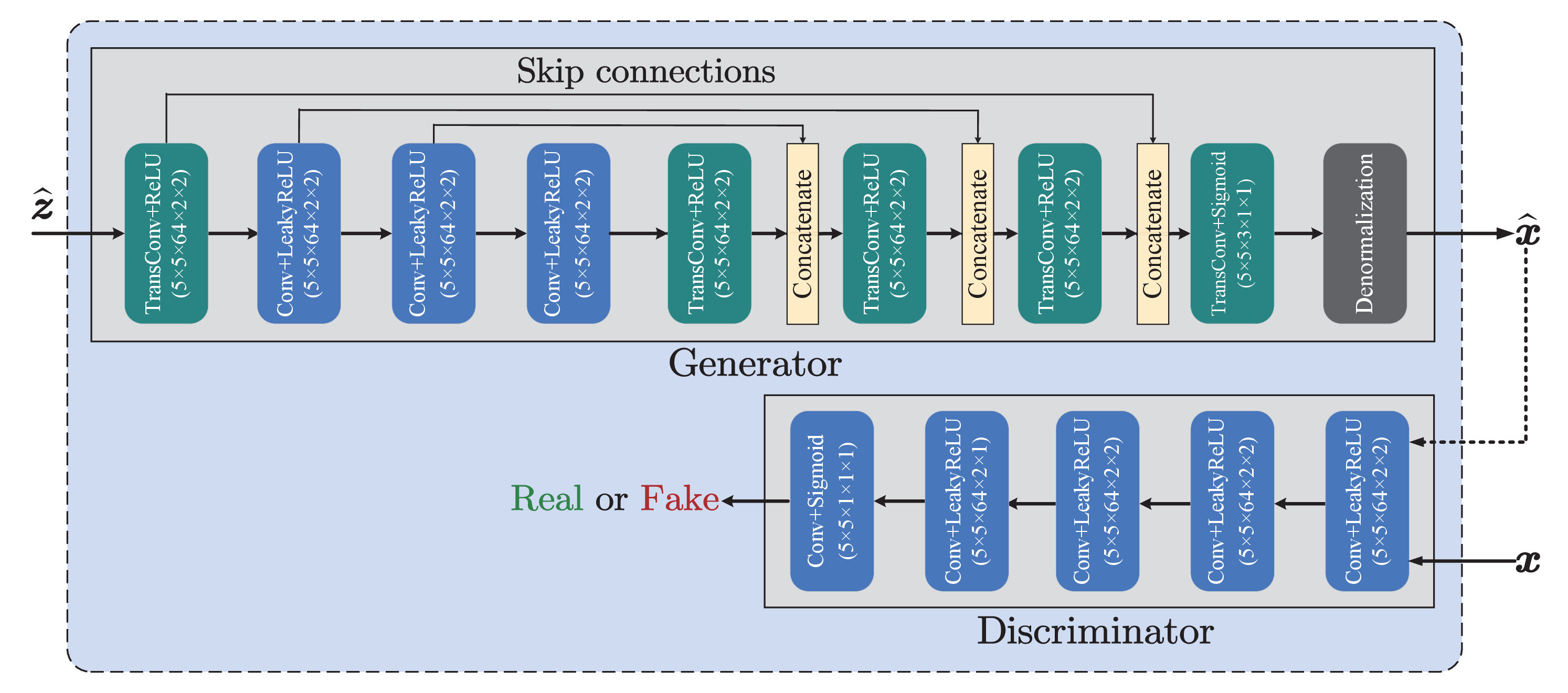}
	\caption{Architecture of the adversarially trained decoder in cGAN-JSCC.}
	\label{fig.3_architecture_decoder_GAN}
\end{figure*}

In the previous subsection, to enhance both the pixel-wise accuracy and the perceptual quality of reconstructed images, we propose G-UNet-JSCC, a DeepJSCC method comprising an encoder and a U-Net-based  decoder. 
Building on this, we propose the cGAN-JSCC method, which is based on a cGAN framework inspired by the Pix2Pix architecture \cite{b25}. In this design, we keep the encoder unchanged from that of G-UNet-JSCC but adversarially train the decoder (i.e., the U-Net-based generator) against a patch-based discriminator. 
Unlike standard GANs for random generation, the cGAN-JSCC decoder takes the noisy and distorted channel output from the encoder as its conditional input and learns to produce more accurate reconstructions via adversarial training, thereby further improving reconstruction performance.
The cGAN-JSCC method employs a two-stage training procedure. In the outer training stage, the encoder and the decoder are trained end-to-end using a weighted sum of the MSE and SSIM losses.
In the inner training stage, we keep the encoder fixed and fine-tune the decoder by introducing adversarial training. Specifically, the U-Net-based generator and the discriminator are jointly optimized using a composite loss that combines the adversarial loss and the L2 loss.


As shown in Fig. \ref{fig.3_architecture_decoder_GAN}, the adversarially trained decoder in cGAN-JSCC takes the corrupted signal $\hat{\boldsymbol{z}}$ as its input. At the receiver, the signal is decoded by the decoder to reconstruct the source image.   
Therefore, the image reconstructed by the decoder can be expressed as
\begin{equation}
	\hat{\boldsymbol{x}} = g_{\boldsymbol{\psi}}\left( \delta\left( f_{\boldsymbol{\theta}}(\boldsymbol{x}), \sigma^2 \right) \right).
\end{equation}   
where $g_{\boldsymbol{\psi}}$	 denotes the decoding function  parameterized by $\boldsymbol{\psi}$ for cGAN-JSCC. 
Subsequently, both the reconstructed image $\hat{\boldsymbol{x}}$ and the original image $\boldsymbol{x}$ are fed into the discriminator. In the decoder, the U-Net-based generator and the discriminator are denoted as $G$ and $D$, respectively. Thus, we also have $\hat{\boldsymbol{x}} = G(\hat{\boldsymbol{z}})$.
During adversarial training, the U-Net-based generator learns to produce realistic image reconstructions that deceive the discriminator, while the discriminator strives to distinguish between the original image and the reconstructed one.


Unlike a regular discriminator that outputs a scalar value, we adopt a patch-based discriminator architecture \cite{b34, b35}. 
In this architecture, the discriminator maps the input to a matrix of responses, where each element corresponds to a distinct receptive field (or patch) and represents a local real/fake classification score.
During training, our discriminator learns to classify each local patch of the input as either real (from $\boldsymbol{x}$) or fake (from $G\left( \hat{\boldsymbol{z}}\right)$).
Subsequently, the patch-based discriminator averages these local scores to produce a final decision on whether the input image is real or fake. 
This patch-level feedback mechanism enables the patch-based discriminator to guide the generator in improving its reconstructions more effectively than a regular discriminator.
Our discriminator comprises five convolutional layers. The first four layers employ the LeakyReLU activation function, while the final layer uses the Sigmoid activation function to scale the output values to the range [0, 1].

The discriminator outputs are denoted as $D_{\text{real}} = D(\boldsymbol{x})$ for real images and $D_{\text{fake}} = D(G\left( \hat{\boldsymbol{z}}\right))$ for reconstructed images produced by the U-Net-based generator.
Each element in $D_{\text{real}}$ or $D_{\text{fake}}$ represents the probability that the corresponding patch of the input is real. 
This matrix of probabilities ranging between 0 and 1 provides detailed feedback to the generator, guiding it to produce more realistic image reconstructions.
Specifically, each element in $D_{\text{real}}$ corresponds to the probability assigned by the discriminator that a specific patch of $\boldsymbol{x}$ is real. Conversely, each element in $D_{\text{fake}}$ indicates the probability assigned by the discriminator that the corresponding patch of $G\left( \hat{\boldsymbol{z}}\right)$ is real (i.e., the generator successfully deceives the discriminator).
During adversarial training, the patch-based discriminator learns to push each element of $D_{\text{real}}$ close to 1, while the U-Net-based generator is trained to make each element in $D_{\text{fake}}$ approach 1. 
Thus, within the proposed decoder, the U-Net-based generator and the patch-based discriminator are trained adversarially.



\begin{algorithm}
	\caption{cGAN-JSCC Joint Training Algorithm}	
	
	\begin{algorithmic}[1]
\State \textbf{Input:} Training dataset $\mathcal{D} = \{\boldsymbol{x}_i\}_{i=1}^T$, the number of epochs $E$, BCR set $R$, batch size $B$, SNR set $S$, learning rate $\eta$


		\State \textbf{Output:} Trained encoders $f_{\boldsymbol{\theta}}$ and decoders $g_{\boldsymbol{\psi}}$ for all $(r,\gamma)$ combinations


		\For{each bandwidth ratio $r \in R$}	
		\For{each SNR value $\gamma \in S$}

		\State Initialize encoder $f_{\boldsymbol{\theta }}$ and decoder $g_{\boldsymbol{\psi}}$

		\State Compute noise variance: $\sigma^2 \gets 10^{-\gamma/10}$	
		
		\For{epoch $= 1$ to $E$}
		
		\State \textbf{\textit{Forward pass through JSCC pipeline}}  
		
		
		\State Sample a minibatch of size $B$ from $\mathcal{D}$		
		

		\State Encode the source image $\boldsymbol{x}$ to the complex-valued symbol vector $\boldsymbol{z}$ using Eq.~\eqref{encoding}: $\boldsymbol{z}\gets f_{\boldsymbol{\theta }}(\boldsymbol{x})$

		
\State Add the AWGN to the vector $\boldsymbol{z}$ using Eq. \eqref{transfer_function} and obtain the corrupted signal $\hat{\boldsymbol{z}}$ 

\State \textbf{\textit{GAN forward pass}}

\State Feed $\hat{\boldsymbol{z}}$ to the generator $G$ to reconstruct the image: $\hat{\boldsymbol{x}} \gets G(\hat{\boldsymbol{z}})$
\State Compute discriminator outputs for real and reconstructed images: $D_{\text{real}} \gets D(\boldsymbol{x})$, $D_{\text{fake}} \gets D(\hat{\boldsymbol{x}})$


\State \textbf{\textit{Compute losses}}

\State Compute the loss $\mathcal{L} _{\mathrm{C}}$ according to Eq.~\eqref{G-UNet-jscc_combined_loss}

\State Compute the generator loss $\mathcal{L}_{G2}$ using Eq.~\eqref{Gen2_loss}

\State Compute the discriminator loss $\mathcal{L}_{D}$ according to Eq.~\eqref{eq:GAN_loss}

\State \textbf{\textit{Joint parameter update via backpropagation}}



\State Update parameters $\{\boldsymbol{\theta}, \boldsymbol{\psi}\}$ by gradient descent on $\mathcal{L} _{\mathrm{C}}$: $\boldsymbol{\theta}, \boldsymbol{\psi} \leftarrow \boldsymbol{\theta}, \boldsymbol{\psi} - \eta \nabla_{\boldsymbol{\theta}, \boldsymbol{\psi}} \mathcal{L}_{\mathrm{C}}$

\State Update generator parameters $\boldsymbol{\varphi}_G$ by descending the gradient of  $\mathcal{L}_{G2}$: $\boldsymbol{\varphi}_G \gets \boldsymbol{\varphi}_G - \eta \nabla_{\boldsymbol{\varphi}_G} \mathcal{L}_{G2}$


\State Update discriminator parameters $\boldsymbol{\varphi }_D$ by gradient descent on $\mathcal{L}_{D}$: $\boldsymbol{\varphi}_D \gets \boldsymbol{\varphi}_D - \eta \nabla_{\boldsymbol{\varphi}_D} \mathcal{L}_{D}$


\EndFor
		
\State Save trained models for current $(r, \gamma)$ configuration: encoder $f_{\boldsymbol{\theta}}$ and decoder $g_{\boldsymbol{\psi}}$

		\EndFor
		\EndFor
		

		
	\end{algorithmic}
	\label{GAN-jscc_training2}	
\end{algorithm}


The core of the proposed cGAN-JSCC method lies in the adversarial training between the U-Net-based generator and the patch-based discriminator, which minimizes an adversarial  loss. The minimization of this loss has been demonstrated to effectively capture both global semantics and local textures, producing visually appealing generated images \cite{b39, b40}. 
The adversarial loss, referred to as the GAN loss,  can be formulated as
\begin{equation}
	\begin{aligned}
		\mathcal{L}_{\mathrm{GAN}}(D, G) &= \mathbb{E}_{\boldsymbol{x}} \left[ \log D(\boldsymbol{x}) \right] \\
		&\quad + \mathbb{E}_{\hat{\boldsymbol{z}}} \left[ \log \left( 1 - D(G(\hat{\boldsymbol{z}})) \right) \right].
	\end{aligned}
	\label{eq:GAN_loss}
\end{equation}
During training, the discriminator aims to maximize $\mathcal{L}_{\mathrm{GAN}}$ (by correctly classifying real and fake images), while the generator aims to minimize it (by fooling the discriminator).  
The loss function for training the generator can be expressed as 
\begin{equation}
	\mathcal{L}_G(D,G) = \mathbb{E}_{\hat{\boldsymbol{z}}}\left[ \log \left( 1-D(G(\hat{\boldsymbol{z}})) \right) \right].
	\label{eq:Gen_loss}
\end{equation}

	


To further improve the quality of the reconstructed images, we incorporate an L2 loss $\mathcal{L}_{\mathrm{L2}}$ \cite{b25, b38} into the generator's loss function. The L2 loss is defined as:
\begin{equation}
	\mathcal{L}_{\mathrm{L}2}\left(G\right)
	=\lambda_2
	\mathbb{E}_{\hat{\boldsymbol{z}},\boldsymbol{x}}
	\left[
	\left\|
	\boldsymbol{x}-G(\hat{\boldsymbol{z}})
	\right\|_2^2
	\right],
	\label{eq:L1_loss}
\end{equation} 
where $\lambda_2$ is the weighting coefficient for the term $\mathbb{E}_{\hat{\boldsymbol{z}},\boldsymbol{x}}
\left[
\left\|
\boldsymbol{x}-G(\hat{\boldsymbol{z}})
\right\|_2^2
\right]$.
Then, the loss defined in Eq.~\eqref{eq:Gen_loss} for training the generator can be rewritten as
\begin{equation}
	\begin{aligned}
		\mathcal{L}_{G2}(D,G) &= \mathbb{E}_{\hat{\boldsymbol{z}}}\left[ \log \left( 1-D(G(\hat{\boldsymbol{z}})) \right) \right] \\
		&\quad + \lambda_2
		\mathbb{E}_{\hat{\boldsymbol{z}},\boldsymbol{x}}
		\left[
		\left\|
		\boldsymbol{x}-G(\hat{\boldsymbol{z}})
		\right\|_2^2
		\right].
		\label{Gen2_loss}
	\end{aligned}
\end{equation}

Thus, the overall adversarial training objective for the decoder in cGAN-JSCC is given by the following minimax optimization problem:
\begin{equation}
	\underset{\boldsymbol{\varphi }_G}{\min}\,\underset{\boldsymbol{\varphi }_D}{\max}\left[ \mathcal{L} _{\mathrm{GAN}}(D,G)+\mathcal{L} _{\mathrm{L}2}\left( G \right) \right],
	\label{eq:total_minmax_loss}
\end{equation}
where $\boldsymbol{\varphi}_G$ and $\boldsymbol{\varphi}_D$ denote the parameter sets of the U-Net-based generator and the discriminator, respectively. The decoder parameters $\boldsymbol{\psi}$ consist of the generator parameters $\boldsymbol{\varphi}_G$ and the discriminator parameters $\boldsymbol{\varphi}_D$.

The total loss can be expressed as $\mathcal{L}_{\mathrm{total}} =\mathcal{L}_{\mathrm{GAN}}(D, G) +\mathcal{L}_{\mathrm{L2}}(G)$.  To optimize this objective, the U-Net-based generator and the discriminator within the adversarially trained decoder are trained alternately to address the minimax problem. 
In each iteration, the parameters of one network are fixed while those of the other are updated.
Specifically, during the generator's update phase, the discriminator parameters $\boldsymbol{\varphi}_D$ are fixed, and the generator aims to minimize the total loss to fool the discriminator. During the discriminator's update phase, $\boldsymbol{\varphi}_G$ is fixed, and the discriminator aims to maximize the adversarial loss $\mathcal{L}_{\mathrm{GAN}}(D, G)$ to better distinguish the source images from the reconstructed ones.
This adversarial training process enhances the cGAN-JSCC method's robustness to diverse channel conditions and drives the generator to produce more realistic and accurate reconstructed images.  
Consequently, the optimal generator $G^*$ is obtained by solving the following minimax optimization problem:
\begin{equation}
	G^* = \arg\min_G \max_D \, \bigl[ \mathcal{L}_{\mathrm{GAN}}(D, G) + \mathcal{L}_{\mathrm{L2}}(G) \bigr].
	\label{eq:minmax_loss}
\end{equation}

\begin{figure}
	\centering
	\includegraphics[width=8.5cm]{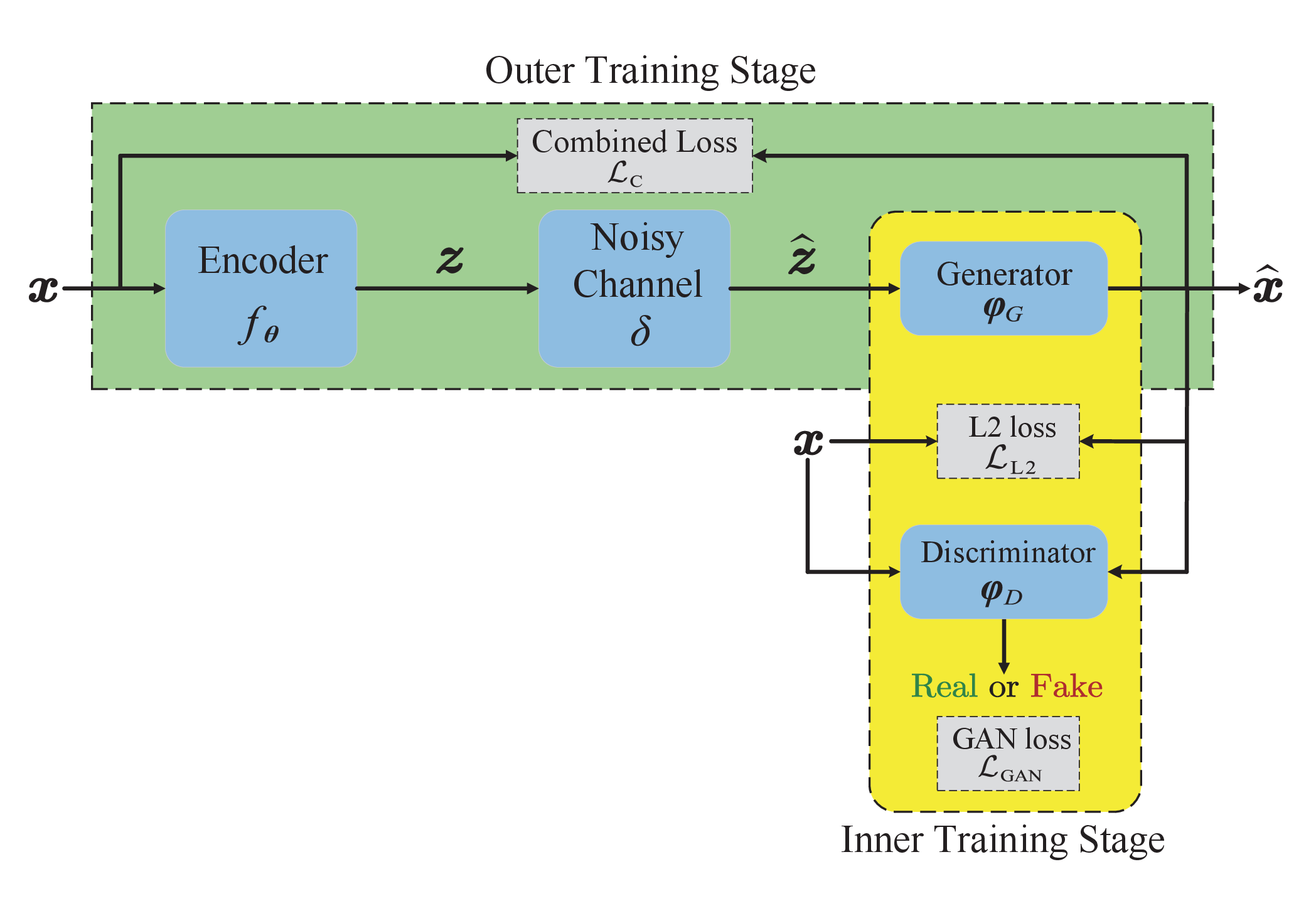}
	\caption{Two-stage training procedure for the cGAN-JSCC scheme.}
	\label{fig4_training_process_GANJSCC}
\end{figure}

The training procedure of cGAN-JSCC is summarized in Algorithm \ref{GAN-jscc_training2}, which
uses a two-stage training strategy.  In the outer training stage, the encoder and the decoder are jointly trained end-to-end using the combined loss defined in Eq.~\eqref{G-UNet-jscc_combined_loss} to achieve high reconstruction performance.
In the inner training stage, the U-Net-based generator and the patch-based discriminator within the decoder are adversarially trained by minimizing a joint loss that combines the adversarial loss and the L2 loss.
The inner training stage further improves reconstruction quality through adversarial training.

Fig. \ref{fig4_training_process_GANJSCC} illustrates the two-stage training architecture of cGAN-JSCC. The green area denotes the outer training stage that performs end-to-end training of the entire JSCC model, while the yellow area corresponds to the inner training stage performing adversarial training.
Specifically, in the outer stage, the encoder and the decoder are jointly optimized end-to-end using a weighted sum of the MSE and SSIM losses. 
In the inner stage, adversarial training is performed within the decoder's forward pass, with the generator and the discriminator jointly trained to minimize a joint loss combining the adversarial and L2 losses.
Similar to  G-UNet-JSCC,  cGAN-JSCC is also particularly suitable for practical deployment scenarios with resource-constrained encoders. It efficiently pairs a computationally-limited encoder with a powerful decoder, thereby achieving high reconstruction quality while maintaining relatively low computational overhead at the transmitter.
During training, cGAN-JSCC undergoes both the outer and inner training stages. 
However, during the deployment (or inference) phase, only the encoder and the U-Net-based generator are used, while the discriminator used during adversarial training is not involved in the inference process.

\begin{figure*}[htbp]  
	\begin{minipage}[t]{0.49\textwidth}
		
		\centering	
		\includegraphics[width=7.5cm]{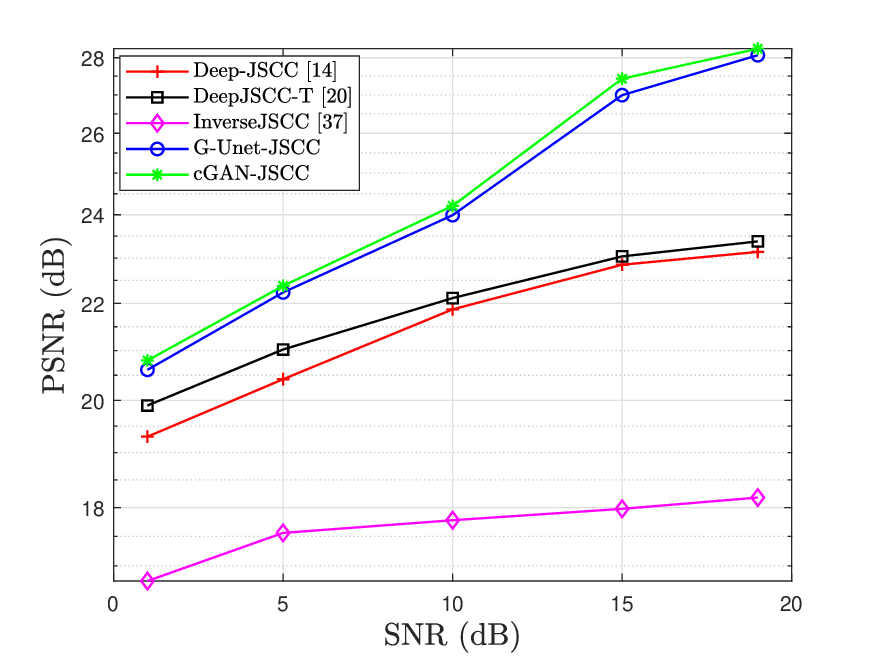} 
		\subcaption{}
	\end{minipage}	
	\begin{minipage}[t]{0.49\textwidth}			
		\centering	
		\includegraphics[width=7.5cm]{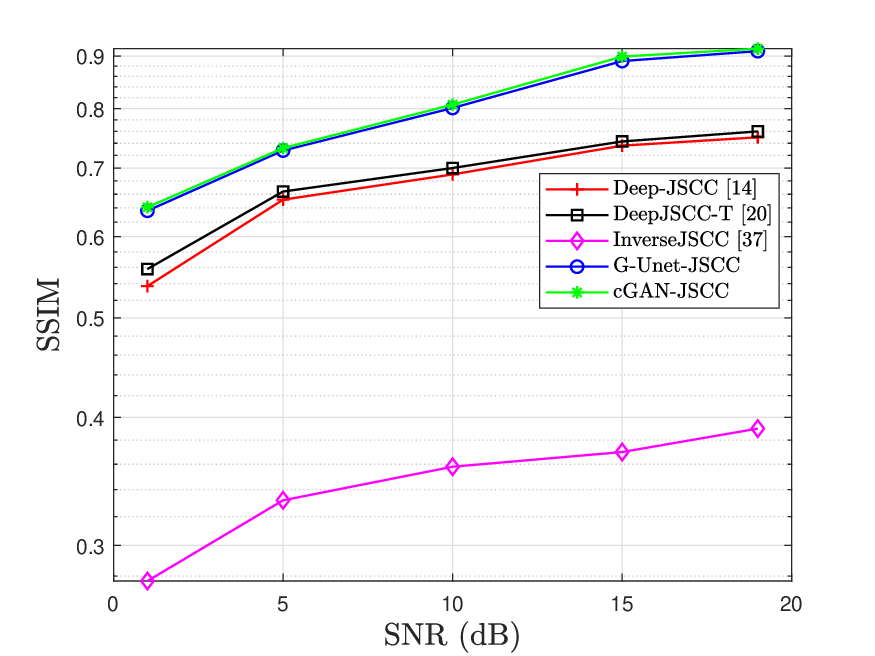} 
		\subcaption{}
	\end{minipage}	
	
	\caption{Performance comparison of different JSCC reconstruction methods on the CIFAR-10 dataset over Rayleigh fading channels at a compression ratio of $r = 1/12$: (a) PSNR and (b) SSIM versus channel SNR.}	
	
	\label{tex9_cifar10_psnr/ssim}		
\end{figure*}

\section{Numerical Results}


This section presents the experimental results comparing the performance of the proposed G-UNet-JSCC and cGAN-JSCC methods. Both DeepJSCC methods were trained and tested over an AWGN channel, as defined in Eq. \eqref{transfer_function}. We conducted experiments on low-resolution images from the CIFAR-10 dataset \cite{b41}, using 50000 images for training and 10000 images for testing. All images in this dataset have a fixed size of $32\times32\times3$. 
For mid-resolution and high-resolution images, we used the CelebA dataset \cite{b42} and the CelebA-HQ dataset \cite{b37}, respectively.
The face images from the CelebA and CelebA-HQ datasets were center-cropped and then resized to $64\times64\times3$ and $256\times256\times3$, respectively.
The CelebA dataset contains 160000 training images and 10000 test images, while  the CelebA-HQ dataset comprises 24000 images for training and 2000  images for testing. 
To mitigate memory constraints, the batch size was set to $B=1$ for training both cGAN-JSCC and G-UNet-JSCC on the aforementioned datasets. 
Unless otherwise stated, all models were trained for $E=20$ epochs.
All encoder and decoder networks were optimized using the Adam optimizer with an initial learning rate of $\eta = 0.0001$. To ensure a fair comparison, each model is independently trained for a specific SNR level and subsequently evaluated under the same SNR condition.  
We implemented the G-UNet-JSCC and cGAN-JSCC methods in Python 3.8.20 with TensorFlow 2.6.0.  All experiments were performed on a workstation with an Intel Xeon Gold 6348 CPU, 383 GB of RAM, and an NVIDIA RTX A6000 GPU.






To ensure that the target BCR $r$ matches the actual BCR, the parameters $k$ and $c$ are determined as follows for all networks during both training and inference.
Let $n = H_{\mathrm{I}}\times W_{\mathrm{I}}\times C_{\mathrm{I}}$, where $H_{\mathrm{I}}$, $W_{\mathrm{I}}$, and $C_{\mathrm{I}}$ represent the height, width, and number of channels of the source image, respectively.  
Then, we set $k=\lfloor r\times n \rfloor$ and $c=\lfloor k / \left( H_{\mathrm{O}}\times W_{\mathrm{O}}/2 \right) \rfloor$, where $\lfloor \cdot \rfloor$ denotes the floor function. Here, $H_{\mathrm{O}}$ and $W_{\mathrm{O}}$ denote the height and width of the feature map output by the last convolutional layer of the encoder, respectively.



\subsection{Evaluation Metrics}



To comprehensively evaluate the quality of JSCC reconstructions, we employ several performance metrics: PSNR for pixel-level fidelity, SSIM for structural similarity, and LPIPS for perceptual quality.

The PSNR is given by
\begin{equation}
	\mathrm{PSNR}(\boldsymbol{x}, \hat{\boldsymbol{x}}) = 10 \log_{10} \left( \frac{P_{\mathrm{MAX}}^2}{\mathrm{MSE}(\boldsymbol{x}, \hat{\boldsymbol{x}})} \right),
\end{equation}
where $P_{\mathrm{MAX}}=255$ denotes the maximum pixel value, and $\mathrm{MSE}(\boldsymbol{x}, \hat{\boldsymbol{x}})$ is the MSE between the original image $\boldsymbol{x}$ and its reconstruction $\hat{\boldsymbol{x}}$.

The SSIM between images $\boldsymbol{x}$ and $\hat{\boldsymbol{x}}$ is defined as
\begin{equation}
	\mathrm{SSIM}(\boldsymbol{x}, \hat{\boldsymbol{x}}) = [l_u(\boldsymbol{x}, \hat{\boldsymbol{x}})]^{\alpha} [c(\boldsymbol{x}, \hat{\boldsymbol{x}})]^{\beta} [s(\boldsymbol{x}, \hat{\boldsymbol{x}})]^{\gamma},
\end{equation}
where $l_u(\cdot)$, $c(\cdot)$, and $s(\cdot)$ measure the luminance, contrast, and structure similarity, respectively:
\begin{align}
l_u(\boldsymbol{x}, \hat{\boldsymbol{x}}) &= \frac{2 \mu_{\boldsymbol{x}} \mu_{\hat{\boldsymbol{x}}} + C_1}{\mu_{\boldsymbol{x}}^2 + \mu_{\hat{\boldsymbol{x}}}^2 + C_1}, \\
	c(\boldsymbol{x}, \hat{\boldsymbol{x}}) &= \frac{2 \sigma_{\boldsymbol{x}} \sigma_{\hat{\boldsymbol{x}}} + C_2}{\sigma_{\boldsymbol{x}}^2 + \sigma_{\hat{\boldsymbol{x}}}^2 + C_2}, \\
	s(\boldsymbol{x}, \hat{\boldsymbol{x}}) &= \frac{\sigma_{\boldsymbol{x}\hat{\boldsymbol{x}}} + C_3}{\sigma_{\boldsymbol{x}} \sigma_{\hat{\boldsymbol{x}}} + C_3}.
\end{align}
Here, $\sigma_{\boldsymbol{x}\hat{\boldsymbol{x}}}$ denotes the sample covariance between $\boldsymbol{x}$ and $\hat{\boldsymbol{x}}$. The standard deviation and the mean of $\boldsymbol{x}$ are denoted by $\sigma_{\boldsymbol{x}}$ and $\mu_{\boldsymbol{x}}$, respectively. Similarly, $\sigma_{\hat{\boldsymbol{x}}}$ and $\mu_{\hat{\boldsymbol{x}}}$ denote the standard deviation and the mean of $\hat{\boldsymbol{x}}$, respectively.
The constants $C_1$, $C_2$, and $C_3$ are included for numerical stability. All parameters ($\alpha$, $\beta$, $\gamma$, $C_1$, $C_2$, $C_3$) are set to their default values following the original implementation in \cite{b43}.


The LPIPS metric was proposed in \cite{b31} to measure the perceptual quality between a source image and its reconstruction.
This metric demonstrates superior alignment with human perception compared to conventional metrics like PSNR and SSIM. The metric works by computing the similarity between the feature activations of two image patches extracted from a pre-trained deep network (e.g., VGG or AlexNet). In this work, we employ the pre-trained VGG network. Lower LPIPS values indicate greater perceptual similarity between the images.

\subsection{Performance Evaluation of Proposed DeepJSCC Methods}

\begin{figure}
	\centering	 
	\includegraphics[width=7.5cm]{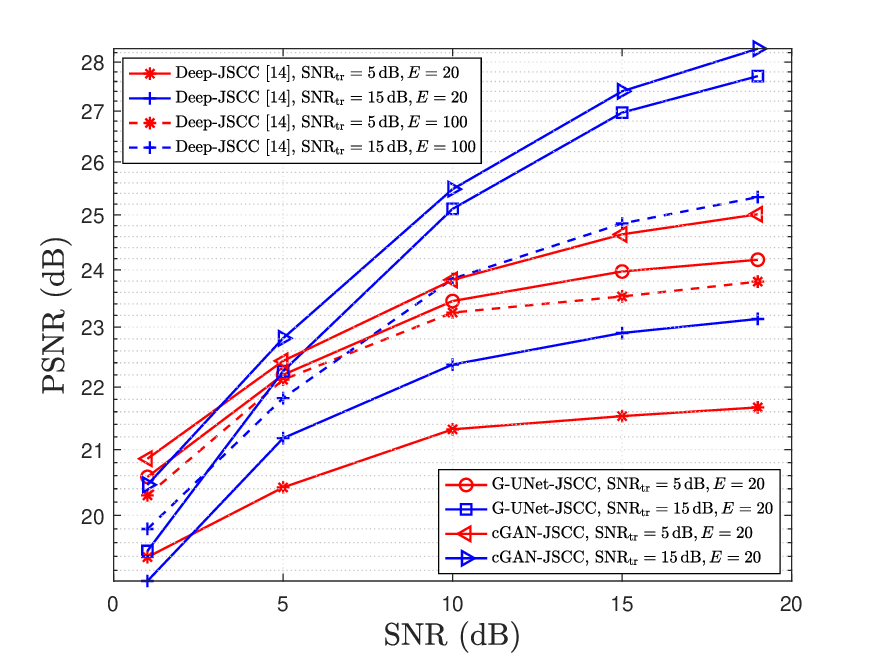} 
	\caption{Robustness of G-UNet-JSCC and cGAN-JSCC to SNR mismatch on the CIFAR-10 dataset over Rayleigh fading channels at $r = 1/12$.  Each model is trained at a fixed SNR and evaluated over a wide range of test SNRs. }
	\label{tex9_cifar10_psnr_variations}  
\end{figure}

This subsection compares the two proposed DeepJSCC schemes, G-UNet-JSCC and cGAN-JSCC, with three existing methods: Deep-JSCC \cite{b11}, DeepJSCC-T \cite{ba17}, and Inverse-JSCC \cite{b30}. 
To comprehensively evaluate the reconstruction quality, we employ three performance metrics: PSNR, SSIM, and LPIPS.
The evaluation is conducted on three image datasets with different image resolutions: CIFAR-10, CelebA, and CelebA-HQ.
For these datasets with different image resolutions, DeepJSCC-T uses the same network architecture.
In this work, we use the term Inverse-JSCC to denote the Inverse-JSCC framework from \cite{b30} without its pre-trained generator. Our implementation of the Inverse-JSCC model corresponds to the forward operator $A$ in \cite{b30} and is trained end-to-end using an MSE loss.
Furthermore, we made several modifications to the original Inverse-JSCC architecture. First, the number of filters in all convolutional layers was reduced from 512 to 256. Second, to ensure compatibility, we inserted a convolutional layer before the decoder's first attention module. This aligns the channel dimension of the reshaped encoder output with that of the first attention module in the Inverse-JSCC decoder. Finally, an additional convolutional layer was appended after the last attention feature (AF) module to reduce the number of output channels to three for RGB reconstruction.

\begin{figure}
	\centering
	\includegraphics[width=8.5cm]{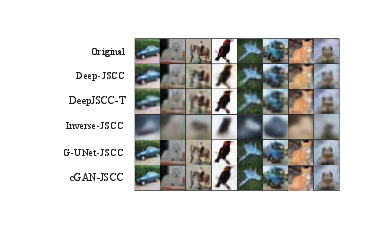}
	\vspace*{-18pt}
	\caption{Visual comparison of JSCC reconstruction methods on the CIFAR-10 dataset over Rayleigh fading channels at SNR = 10 dB and $r = 1/12$. The first row shows the original images, while the second to fifth rows present the reconstructions obtained by Deep-JSCC, DeepJSCC-T, Inverse-JSCC, G-UNet-JSCC, and cGAN-JSCC, respectively.}
	\label{fig.7_Reconstructions_cifar10_10dB}
\end{figure}

\begin{figure*}[htbp]
	\centering 
	\begin{minipage}[t]{0.32\textwidth} 
		\centering
		\includegraphics[width=6.0cm]{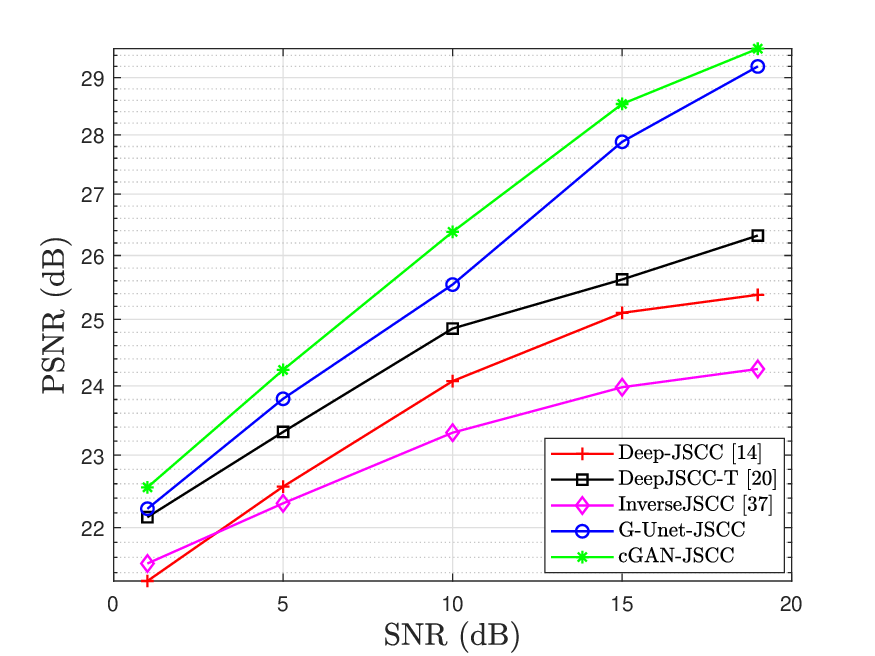} 
		\subcaption{PSNR versus SNR.}
	\end{minipage}
	\hspace{0.72em}
	\begin{minipage}[t]{0.32\textwidth} 
		\centering
		\includegraphics[width=6.0cm]{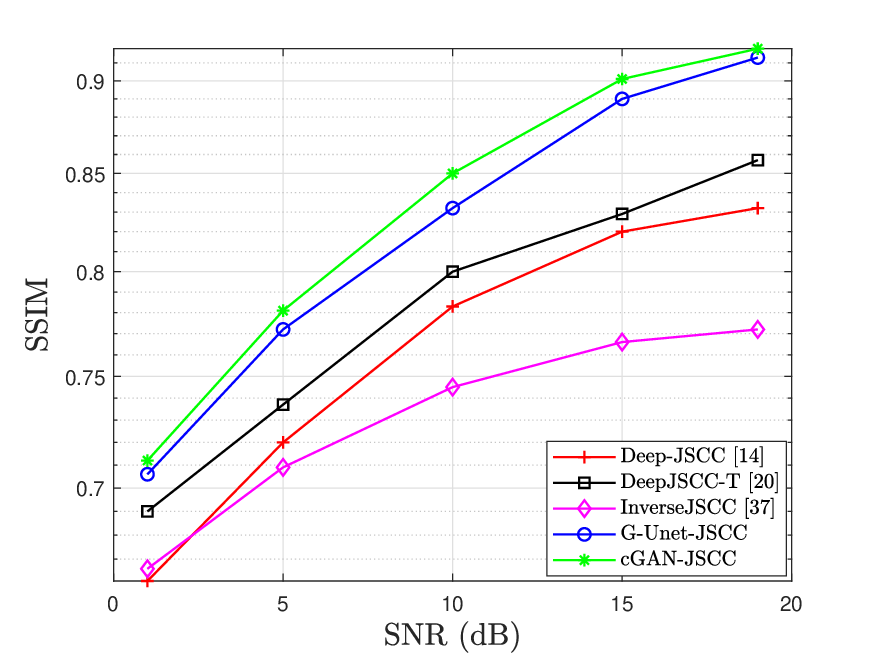}
		\subcaption{SSIM versus SNR.}
	\end{minipage}
	\hspace{0.72em}
	\begin{minipage}[t]{0.32\textwidth} 
		\centering
		\includegraphics[width=6.0cm]{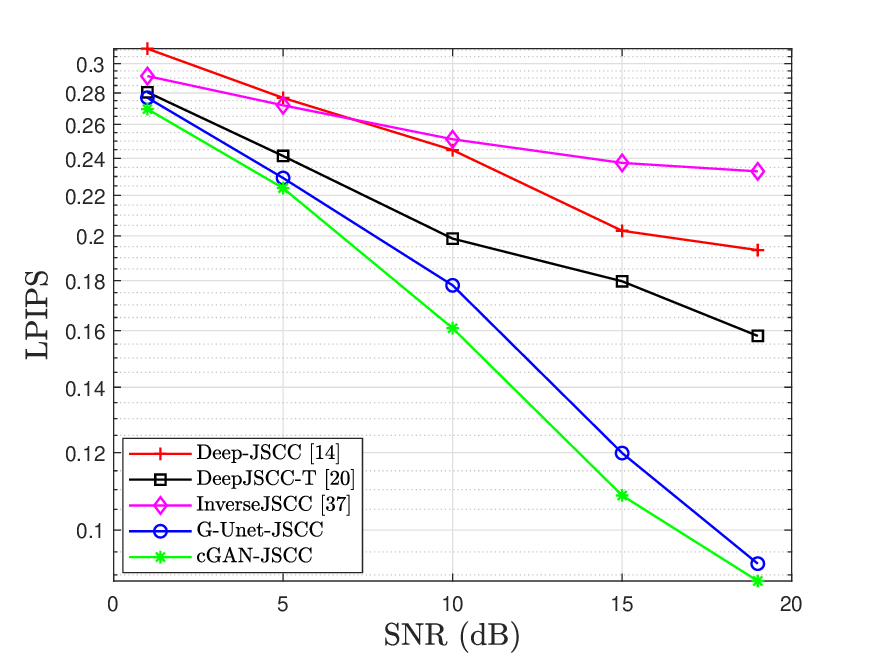}
		\subcaption{LPIPS versus SNR.}
	\end{minipage}
	
	\caption{Performance comparison of different JSCC reconstruction methods on the CelebA dataset over Rayleigh fading channels at $r = 1/12$: (a) PSNR, (b) SSIM, and (c) LPIPS versus channel SNR.}
	\label{tex9_celebA_fig8}
		\vspace*{-18pt}
\end{figure*}

\begin{figure*}  
	\centering
	\includegraphics[width=12.5cm]{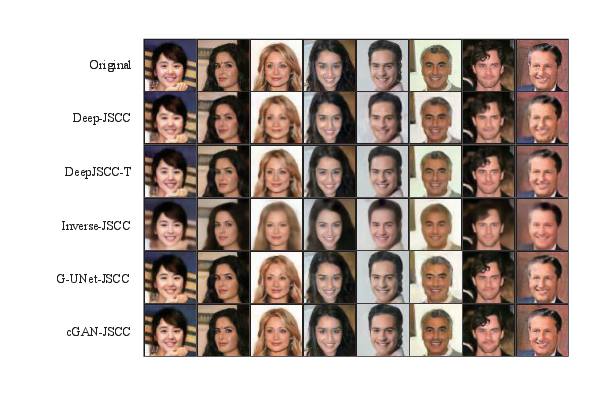} 
	\vspace*{-10pt}
	\caption{Visual comparison of JSCC reconstruction methods on the CelebA dataset over Rayleigh fading channels at SNR = 10 dB and $r = 1/12$. The first row shows the original images. Reconstructions by Deep-JSCC, DeepJSCC-T, Inverse-JSCC, G-UNet-JSCC, and cGAN-JSCC are shown in rows 2 to 5, respectively.}
	\label{fig.9_Reconstructions_celebA_10dB}
\end{figure*}

Fig.~\ref{tex9_cifar10_psnr/ssim} compares the PSNR and SSIM performance of various DeepJSCC methods over Rayleigh fading channels.
As shown in Fig. \ref{tex9_cifar10_psnr/ssim}(a), cGAN-JSCC slightly outperforms G-UNet-JSCC across all considered SNR regimes. This result suggests that adversarial training may enable cGAN-JSCC to extract more effective features than G-UNet-JSCC for low-resolution CIFAR-10 images. In addition, G-UNet-JSCC outperforms DeepJSCC-T and Deep-JSCC. DeepJSCC-T performs slightly better than Deep-JSCC, whereas Deep-JSCC significantly outperforms Inverse-JSCC. Therefore, Inverse-JSCC achieves the lowest PSNR among all compared methods in Fig. \ref{tex9_cifar10_psnr/ssim}(a).
This is primarily because the aggressive downsampling used in the Inverse-JSCC encoder produces feature maps with insufficient information content for low-resolution images, thereby limiting the decoder's reconstruction quality. 
Fig.~\ref{tex9_cifar10_psnr/ssim}(b) shows that cGAN-JSCC achieves the best SSIM performance among all compared methods. G-UNet-JSCC achieves SSIM performance comparable to that of cGAN-JSCC, significantly outperforming both DeepJSCC-T and Deep-JSCC. DeepJSCC-T slightly surpasses Deep-JSCC, whereas Deep-JSCC significantly outperforms Inverse-JSCC. Overall, cGAN-JSCC achieves slightly better reconstruction performance than G-UNet-JSCC for low-resolution CIFAR-10 images. Inverse-JSCC exhibits limited reconstruction performance on such low-resolution images.

Next, we test the robustness of cGAN-JSCC and G-UNet-JSCC to channel SNR variations in Fig. \ref{tex9_cifar10_psnr_variations}. Each curve is obtained by training an end-to-end model at a fixed channel SNR, denoted as $\mathrm{SNR}_{\mathrm{tr}}$, and then evaluating the reconstruction performance of the trained model on the CIFAR-10 test set under different test SNRs. At $E=20$, all three models trained at $\mathrm{SNR}_{\mathrm{tr}}=15~\mathrm{dB}$ outperform their counterparts trained at $\mathrm{SNR}_{\mathrm{tr}}=5~\mathrm{dB}$ when evaluated at a test SNR of 5 dB. This result may be attributed to insufficient convergence under the 20-epoch training setting. In contrast, at $E=100$, Deep-JSCC with $\mathrm{SNR}_{\mathrm{tr}}=5~\mathrm{dB}$ outperforms its counterpart with $\mathrm{SNR}_{\mathrm{tr}}=15~\mathrm{dB}$ at a test SNR of 5 dB.
Unlike conventional separation-based systems, none of the three models exhibits a pronounced cliff effect when the test SNR falls below $\mathrm{SNR}_{\mathrm{tr}}$. Nevertheless, all three methods exhibit performance degradation as the test SNR decreases. Across the considered test SNR range, both G-UNet-JSCC and cGAN-JSCC achieve higher PSNR values than the Deep-JSCC models with $E=20$ under both training SNR settings. For each training SNR setting, cGAN-JSCC consistently achieves the highest PSNR among the three methods.

We present a visual comparison of different DeepJSCC methods on the CIFAR-10 dataset over Rayleigh fading channels in Fig. \ref{fig.7_Reconstructions_cifar10_10dB}. Among the compared methods, Inverse-JSCC produces the most severely blurred reconstructions. In contrast, the reconstructions from both DeepJSCC-T and Deep-JSCC preserve image colors and details more effectively than those from Inverse-JSCC. Both cGAN-JSCC and G-UNet-JSCC produce visually clearer reconstructions, preserving image colors and details more effectively than DeepJSCC-T and Deep-JSCC.
Moreover, the visual differences between cGAN-JSCC and G-UNet-JSCC are relatively small.

Fig. \ref{tex9_celebA_fig8} compares different DeepJSCC methods in terms of PSNR, SSIM, and LPIPS on the CelebA dataset across a range of channel SNR values. At an SNR of 5 dB, Inverse-JSCC achieves a slightly lower LPIPS than Deep-JSCC, although Deep-JSCC achieves higher PSNR and SSIM than Inverse-JSCC. At higher SNRs, Deep-JSCC outperforms Inverse-JSCC in terms of PSNR, SSIM, and LPIPS.
In terms of PSNR, SSIM, and LPIPS, DeepJSCC-T consistently outperforms Deep-JSCC across the considered SNR range, while G-UNet-JSCC consistently outperforms DeepJSCC-T across the same SNR range. With adversarial training, cGAN-JSCC achieves the best performance among the considered methods in terms of PSNR, SSIM, and LPIPS, consistently outperforming G-UNet-JSCC.




Fig. \ref{fig.9_Reconstructions_celebA_10dB} provides a visual comparison of different DeepJSCC methods on the CelebA dataset. The reconstructions produced by Deep-JSCC are relatively blurry, while those generated by Inverse-JSCC appear visually plausible but fail to preserve some of the color information and fine details of the original images. In addition, the Inverse-JSCC reconstructions exhibit noticeable differences from the corresponding original images. DeepJSCC-T produces visually clearer reconstructions than Deep-JSCC. The reconstructions of G-UNet-JSCC seem to preserve color information and fine details slightly better than those of DeepJSCC-T. cGAN-JSCC appears to achieve better visual reconstruction quality than G-UNet-JSCC.



\begin{figure*}[htbp]
	\centering 
	\begin{minipage}[t]{0.32\textwidth} 
		\centering
		\includegraphics[width=6.0cm]{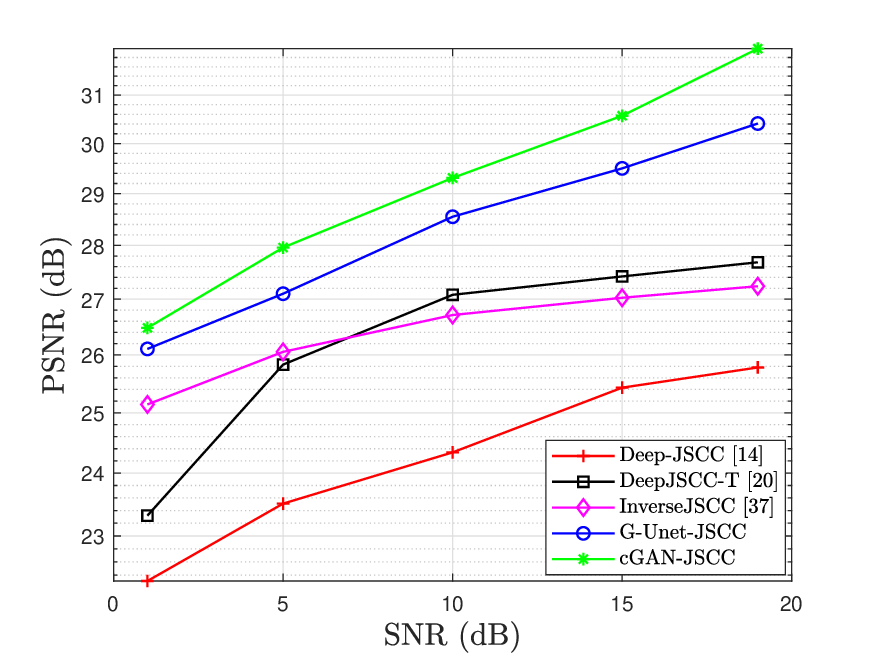} 
		\subcaption{PSNR versus SNR.}
	\end{minipage}
	\hspace{0.72em}
	\begin{minipage}[t]{0.32\textwidth} 
		\centering
		\includegraphics[width=6.0cm]{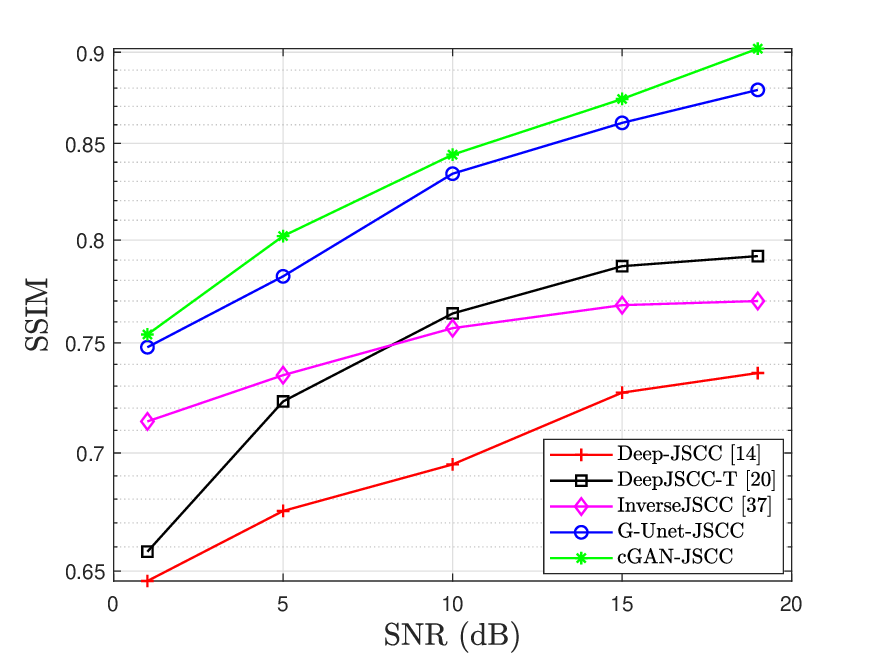}
		\subcaption{SSIM versus SNR.}
	\end{minipage}
	\hspace{0.72em}
	\begin{minipage}[t]{0.32\textwidth} 
		\centering
		\includegraphics[width=6.0cm]{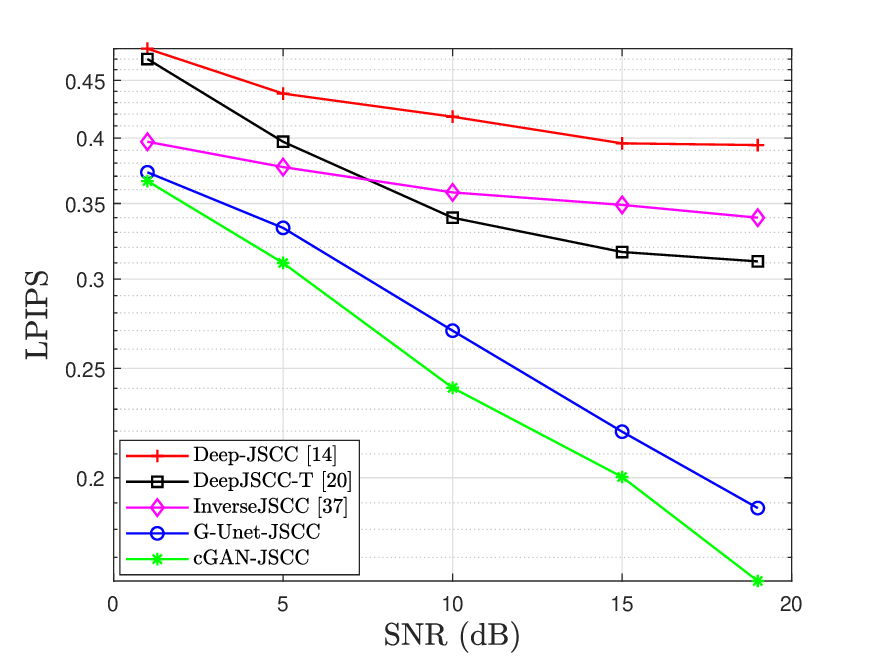}
		\subcaption{LPIPS versus SNR.}
	\end{minipage}
	
	\caption{Performance comparison of different reconstruction methods on the CelebA-HQ dataset over Rayleigh fading channels at $r = 1/12$: (a) PSNR, (b) SSIM, and (c) LPIPS versus channel SNR.}
	\label{tex9_celebA-HQ_fig10}
\end{figure*}

Fig. \ref{tex9_celebA-HQ_fig10} compares the PSNR, SSIM, and LPIPS performance of various DeepJSCC methods on the CelebA-HQ dataset across a range of channel SNR values. 
We observe that Inverse-JSCC outperforms Deep-JSCC in terms of PSNR, SSIM, and LPIPS. This result demonstrates that Inverse-JSCC maintains good performance on the high-resolution facial images from the CelebA-HQ dataset and is consistent with the results reported in \cite{b30}. 
Furthermore, Inverse-JSCC performs better than DeepJSCC-T in terms of PSNR, SSIM, and LPIPS at low SNRs.
G-UNet-JSCC outperforms both Inverse-JSCC and DeepJSCC-T in terms of PSNR, SSIM, and LPIPS across the considered SNR range, while cGAN-JSCC outperforms G-UNet-JSCC in terms of all three metrics over the same SNR range on the CelebA-HQ dataset.



\begin{figure*}
	\centering
	\includegraphics[width=15cm]{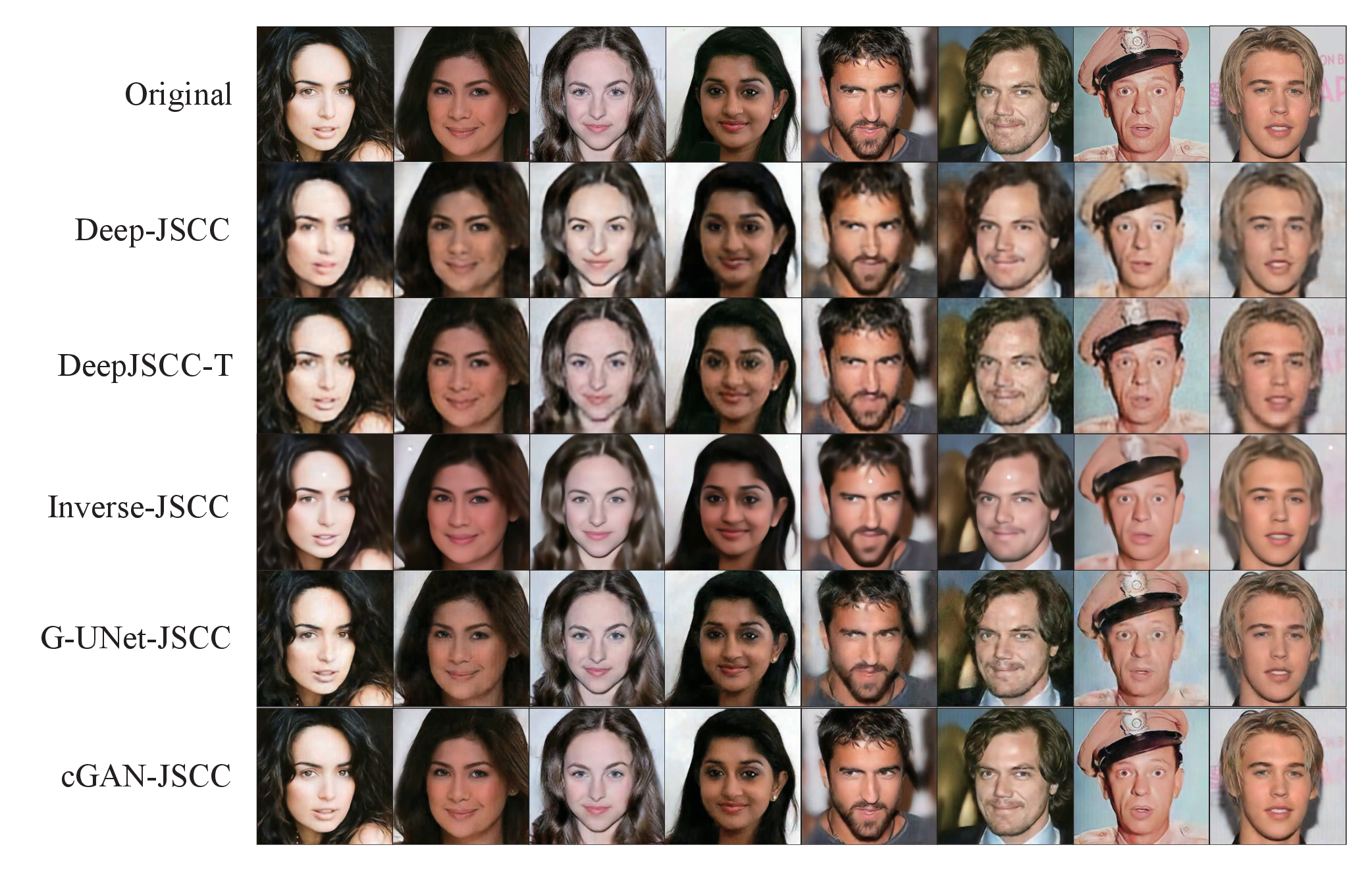}
	\caption{ Visual comparison of JSCC reconstruction methods on the Celeb-HQ dataset over Rayleigh fading channels at SNR = 5 dB and $r = 1/12$. The first row shows the original images, while the second to fifth rows present the reconstructions obtained by Deep-JSCC, DeepJSCC-T, Inverse-JSCC, G-UNet-JSCC, and cGAN-JSCC, respectively.}
	\label{fig11_Reconstructions_celebHQ_5dB}
\end{figure*}

Fig. \ref{fig11_Reconstructions_celebHQ_5dB} presents the visual comparison of different DeepJSCC methods on the CelebA-HQ dataset over Rayleigh fading channels at an SNR of 5 dB.
Although Deep-JSCC reconstructs face images that are not blurry at SNR = 5 dB, it still fails to accurately reproduce the color and fine details of the original high-resolution images.
DeepJSCC-T appears to produce visually clearer reconstructions than Deep-JSCC.
Inverse-JSCC produces reconstructions of high perceptual quality on the CelebA-HQ dataset, which is consistent with the results reported in \cite{b30}.
However, the Inverse-JSCC reconstructions frequently exhibit an isolated white point artifact. 
G-UNet-JSCC not only generates reconstructions of high perceptual quality but also appears to capture fine image details better than Inverse-JSCC, while cGAN-JSCC achieves better visual reconstruction quality than G-UNet-JSCC on the CelebA-HQ dataset.

\begin{table}[htbp]
	\centering
	\caption{Comparison of FLOPs and parameters between the proposed methods and existing DeepJSCC methods.}
	
	\scalebox{1}
	{
		\begin{tabular}{c|c|c}
			
			\toprule[0.8pt]
			Model	& Parameters (K)  &	FLOPs (M) 
			\\   \hline  
			

			Deep-JSCC	& 164.4 &  19.3  \\ \hline  

			DeepJSCC-T	& 4670.2 &  206.8  \\ \hline

			Inverse-JSCC	& 76973.7  & 26384.7  \\ \hline 

G-UNet-JSCC	&   929.0  &	259.7   \\ \hline  

cGAN-JSCC  &  929.0  &	259.7  \\   

			\bottomrule[0.8pt]
		\end{tabular}	
	}
	\label{DeepJSCC_complexity}
	\vspace{-0.5mm}
\end{table}

\subsection{Complexity Analysis}

This subsection analyzes the number of parameters and floating-point operations (FLOPs) of the two proposed DeepJSCC schemes, G-UNet-JSCC and cGAN-JSCC, and three existing DeepJSCC methods: Deep-JSCC, DeepJSCC-T, and Inverse-JSCC. 
The model complexity of these methods is evaluated during inference on the CIFAR-10 dataset using a batch size of 1, in terms of parameter count and inference FLOPs.


For all DeepJSCC methods, the computational complexity is mainly determined by their encoder and decoder networks, which primarily consist of 2D convolutional and transposed convolutional layers. Following the analysis in \cite{b35, b44}, the FLOPs of a convolutional or transposed convolutional layer can be estimated as
\begin{equation}
	C_l \sim \mathcal{O}\left(M_{l,h}M_{l,w}F_l^2K_lK_{l-1}\right),
	\label{complexity_en_de}
\end{equation}
where $F_l$ is the kernel size, $K_l$ and $K_{l-1}$ denote the numbers of output and input feature maps, respectively, and $M_{l,h}\times M_{l,w}$ represents the spatial dimensions of the output feature map at the $l$-th layer. Thus, the total FLOPs of an encoder or decoder with $L$ layers can be obtained by summing $C_l$ over all layers. The total FLOPs of each DeepJSCC method are then obtained by summing the FLOPs of its encoder and decoder.

The FLOPs and parameter count of the DeepJSCC methods are summarized in Table \ref{DeepJSCC_complexity}. The parameter count and FLOPs of cGAN-JSCC are the same as those of G-UNet-JSCC, since the discriminator used for adversarial training is not involved in the inference process. G-UNet-JSCC exhibits higher computational complexity than Deep-JSCC, primarily due to its more complex decoder architecture. For the transposed convolutional layers incorporating skip connections, the number of input feature maps $K_{l-1}$ is doubled due to channel concatenation, which further increases the FLOPs of the decoder.
The substantially larger parameter count of DeepJSCC-T than G-UNet-JSCC can be attributed to the larger numbers of input and output feature maps across the layers of DeepJSCC-T. However, the FLOPs of DeepJSCC-T remain comparable to those of G-UNet-JSCC, as the FLOPs of G-UNet-JSCC are also affected by the spatial dimensions of the output feature maps.
Inverse-JSCC incurs the highest computational cost in terms of both FLOPs and parameter count among all compared methods. Compared with G-UNet-JSCC, Inverse-JSCC uses a larger number of feature maps across its layers, a deeper network structure, and more complex modules such as attention mechanisms and residual blocks.
Overall, G-UNet-JSCC and cGAN-JSCC offer a favorable trade-off between model complexity and reconstruction performance, particularly compared with Inverse-JSCC and DeepJSCC-T.

\section{Conclusion}

In this work, we proposed G-UNet-JSCC and cGAN-JSCC, two DeepJSCC methods based on deep generative architectures for wireless image transmission. G-UNet-JSCC comprises an encoder and a U-Net-based generator serving as the decoder. This architecture enables multi-scale feature fusion by combining low-level features with high-level features, thereby improving both pixel-level fidelity and perceptual quality. To further improve pixel-level fidelity, the encoder and the decoder are jointly optimized using a weighted sum of MSE and SSIM losses. 
Building upon G-UNet-JSCC, we further developed cGAN-JSCC by enhancing its decoder while retaining the same encoder. The decoder's generator is adversarially trained against a patch-based discriminator. cGAN-JSCC uses a two-stage training procedure. The outer stage trains the encoder and the decoder end-to-end using the same weighted sum of MSE and SSIM losses, while the inner stage performs adversarial training of the generator and the discriminator within the decoder by minimizing a joint adversarial and L2 loss.
Extensive numerical experiments demonstrate that the proposed methods achieve high pixel-wise fidelity and perceptual quality on both high- and low-resolution images.   
cGAN-JSCC further enhances the reconstruction performance of G-UNet-JSCC by introducing a discriminator to enable adversarial training.

\vspace{12pt}

\ifCLASSOPTIONcaptionsoff
  \newpage
\fi



%


\end{document}